\PassOptionsToPackage{tbtags}{amsmath}
\documentclass[%
 aip,
 amsmath,amssymb,
 preprint,%
 citeautoscript,
author-numerical,
]{revtex4-1}
\usepackage{CJKutf8}
\usepackage{graphicx}
\usepackage{dcolumn}
\usepackage{bm}

\usepackage[utf8]{inputenc}
\usepackage{subfigure}
\usepackage[T1]{fontenc}
\usepackage{mathptmx}
\usepackage{etoolbox}
\usepackage{verbatim}
\usepackage{overpic}
\usepackage{multirow}
\usepackage{upgreek}
\usepackage{amsmath}
\usepackage{amssymb}

\usepackage[dvipsnames]{xcolor}

\usepackage{dcolumn}
\newcolumntype{d}[1]{D{.}{.}{#1}}

\usepackage{longtable}

\usepackage{threeparttable}

\usepackage{hyperref}
\hypersetup{
    colorlinks=true,            
    linkcolor=blue,             
    filecolor=magenta,          
    urlcolor=cyan,              
    pdfpagemode=FullScreen,
    }

\usepackage{amsthm}
\newtheoremstyle{query}%
{}{}
{\color{black}}
{}
{\sffamily}{:}{12pt}
{}
\theoremstyle{query}
\newtheorem{aq}{Author Query/Comment}

\newcommand{\baq}{\begin{aq}}
\newcommand{\eaq}{\end{aq}}
\graphicspath{{figures/}}
\DeclareGraphicsExtensions{.pdf,.eps,.png,.jpg,.jpeg,.bmp}

\makeatletter
\def\@email#1#2{%
 \endgroup
 \patchcmd{\titleblock@produce}
  {\frontmatter@RRAPformat}
  {\frontmatter@RRAPformat{\produce@RRAP{*#1\href{mailto:#2}{#2}}}\frontmatter@RRAPformat}
  {}{}
}%
\makeatother

\begin{document}
\begin{CJK*}{UTF8}{gbsn}

\preprint{Manuscript}

\title{Utility of High-Order Scheme for Unsteady Flow Simulations: Comparison with Second-Order Tool}
\author{Peng Jiang}
\affiliation{School of Ocean and Civil Engineering, Shanghai Jiao Tong University, Shanghai, 200240, China}

\author{Yichen Huang}
\affiliation{School of Ocean and Civil Engineering, Shanghai Jiao Tong University, Shanghai, 200240, China}

\author{Yong Cao}
\affiliation{School of Ocean and Civil Engineering, Shanghai Jiao Tong University, Shanghai, 200240, China}

\author{Shijun Liao}
\affiliation{School of Ocean and Civil Engineering, Shanghai Jiao Tong University, Shanghai, 200240, China}

\author{Bin Xie}
\email{xie.b.aa@sjtu.edu.cn}
\affiliation{School of Ocean and Civil Engineering, Shanghai Jiao Tong University, Shanghai, 200240, China}

\begin{abstract}
The objective of this work is to investigate the utility and effectiveness of the high-order scheme for simulating unsteady turbulent flows. To achieve it, the studies are conducted from two perspectives: (i) the ability of different numerical schemes for turbulence problems under the same set of meshes; and (ii) the accuracy and stability of higher-order schemes for solving turbulence statistics for different mesh types (hexahedral, tetrahedral, and polyhedral cells). The simulations employ the third-order scheme for spatial discretization of the governing equations, while a widely-used second-order solver, namely pisoFoam, is employed for comparison. This study considers the canonical cases of the Taylor-Green vortex (TGV) problem at $Re=100,\, 1600$ and flow past a sphere at $Re=10\,000$ to address the aforementioned two key issues. For the TGV case, the high-order model significantly improves the numerical accuracy with convergence rates and reduces the numerical dissipation of nearly 1/10 of pisoFoam on different meshing types. In the latter case, the high-order scheme with large-eddy simulation (LES) accurately predicts the vortex structures and the flow instability, regardless of grid type. However, pisoFoam is found to be sensitive to mesh types, which results in numerous non-physical structures in the flow field due to numerical noise rather than flow physics, particularly for tetrahedral cells. Furthermore, for the typical low- and high-order flow statistics, the numerical results predicted by the present model show better agreement with the reference data and have less dependence on the type of grids compared with the conventional scheme. In addition, the obtained energy spectrum by the high-order solver accurately captures the Kelvin-Helmholtz (K-H) instability and the vortex shedding frequency, while these important features are less pronounced by the traditional low-order model.
\end{abstract}

\maketitle
\end{CJK*}
\section{Introduction}
Computational fluid dynamics (CFD) for practical applications necessitates numerical methods that optimally balance solution accuracy and computational cost across a wide range of problems \cite{Wang2013HighOrder, VERMEIRE2017GPUHighOrder}. Additionally, these methods must be adaptable to manage geometric complexity and intricate dynamic processes, such as those encountered in underwater vehicles and aircraft, which are often described using unstructured mixed-type meshes. To meet these requirements, most traditional unstructured CFD numerical methods employ second-order accurate schemes for the spatial discretization of governing equations. Developed primarily between the 1970s and 1990s, these methods include solvers implemented in widely used industrial tools like STAR-CCM+ and Fluent \cite{Wang2013HighOrder}. It has been demonstrated in the past few decades that numerical schemes with second-order accuracy can be an effective tool in addressing steady-state problems, including those solved using the Reynolds Averaged Navier-Stokes (RANS) method. However, second-order finite volume methods (FVM) are known to be inadequate for certain flow problems, especially in computational aeroacoustics, vortex-dominant flow, and turbulent flow. It is reported by Xie \textit{et al.} \cite{xie2017piso} that traditional second-order schemes have excessive numerical dissipation, which degrades solution quality near vortical structures and shock waves. 

Consequently, there has been a notable increase in the less-dissipative and non-oscillatory high-order methods used on unstructured grids with at least third-order accuracy for spatial discretizations \cite{EKATERINARIS2005192}. These methods include the discontinuous Galerkin (DG), spectral volume (SV), spectral difference (SD), flux reconstruction (FR), and other similar approaches. It has been suggested that these higher-order methods can be more accurate for the simulation of unsteady turbulent flows \cite{VERMEIRE2017GPUHighOrder}. Among these high-order methods, FVMS3 (Finite Volume method based on Merged Stencil with 3rd-order reconstruction) proposed by Xie \textit{et al.} \cite{Xie2019High-fidelitysolver}, is particularly appealing as it makes quadratic reconstruction on a merged stencil which significantly improves numerical accuracy without significantly increasing computational cost.
Several direct numerical simulations have been carried out for some benchmark problems and the results show that this third-order scheme accuracy can significantly reduce undesired numerical dissipation and dispersion even after long-duration computations, making it a desirable tool for investigating turbulent flows. Moreover, this method has been applied for the numerical simulation of wave breaking \cite{Jiang2022WaveBreaking}. While high-order models have proven advantageous in direct numerical simulation (DNS) of low and medium Reynolds number turbulence problems, the effectiveness of combining high-order schemes with LES for solving complex high-Reynolds-number turbulence problems remains unclear. 

In addition to the numerical methods mentioned above, the characteristics of the computational mesh, including cell types and arrangements, are critical in determining the accuracy of LES\cite{Wang2021mesh}. These features not only affect discretization accuracies due to mesh quality and mesh types but also contribute to turbulence modeling errors due to the filtering process\cite{Wang2021mesh, Celik2005Index}. Regarding discretization errors caused by mesh types, each cell type has its advantages and limitations, consequently, the areas of application vary. Therefore, to evaluate the capability of different cell types, it is important to make comparative studies based on the results of the numerical simulations. Recently, the most commonly used mesh types have been hexahedral orthogonal meshes and non-orthogonal meshes. Non-orthogonal meshes include tetrahedral meshes and polyhedral meshes. It should be noted that in the case of hexahedral orthogonal meshes, the connectivity between adjacent cells in the hexahedral orthogonal mesh was straightforward, and the cell edges formed continuous grid lines. In contrast, the tetrahedral and polyhedral meshes were characterized by arbitrary cell shapes and the ability to be assembled freely within the computational domain \cite{Wang2021mesh}. For simplified geometries, such as spheres and cylinders, the hexahedral grid has been extensively employed due to its superior accuracy and enhanced convergence. Nonetheless, the application of hexahedral mesh becomes challenging, and in some cases impractical, for complex geometries, such as complex submarine models with appendages. To achieve LES for more complex geometries, it is necessary to use the non-orthogonal mesh which tends to be an important alternative to hexahedral mesh due to its high flexibility and adaptability to varying shapes. The tetrahedral mesh has a well-established history of application in non-orthogonal mesh configurations and has been widely used in numerical simulations for various applications, including wind passing through urban buildings \cite{Saeedi2015, Wang2021mesh} and flow past a sphere\cite{rodriguez2011, RODRIGUEZ2013DNS}. \textcolor{black}{Recently, the popularity of polyhedral mesh has been on the rise since it delivers higher numerical accuracy and efficiency with fewer mesh numbers, and provides more flexibility allowing for better adaptation to intricate shapes and boundaries without compromising mesh quality.  It has been used across various areas, including flow around urban buildings\cite{Wang2021mesh, LI2023polyhedraMesh}, and the buoyancy-driven turbulent cavity flow\cite{Addad2015Cavity}.
} However, most previous investigations of cell type effects on LES have been based on traditional low-order numerical schemes. In particular, the systematic study of the performance of the high-order numerical method with LES using the unstructured mesh types, including the polyhedral mesh and the tetrahedral mesh, presents challenges in predicting turbulent flow characteristics around the basic shape of bluff bodies, which is particularly suitable for studying complex geometries.

The goal of this paper is to study the utility and effectiveness of the high-order scheme for simulating unsteady turbulent flows. To achieve it, the studies were conducted from two perspectives: (i) the ability of different numerical schemes for turbulence problems under the same set of meshes; and (ii) the accuracy and stability of higher-order schemes for solving turbulence statistics for different mesh types of hexahedral, tetrahedral, and polyhedral cells, which is attractive for further application of the high-order method to more complex geometries. These objectives are achieved by comparing the accurate third-order solver with a widely used second-order solver, namely pisoFoam. This study mainly considers two canonical cases which are the Taylor-Green vortex problem at $Re$ = 100, 1\,600 and the turbulent flow around a sphere at $Re$ = 10\,000 to address the aforementioned issues. In the first case, we are interested in quantifying the solution accuracy, convergence behavior, energy conservation, and numerical dissipation. For the latter case, meticulous analysis was conducted of the instantaneous flow structures mean flow topology, and spectrum characteristics.

Below, the paper is organized as follows: Sec.~\ref{sec:numerical_model} gives a brief explanation of the methodology. In particular, a general description of the governing equations, the discretization algorithm, and the numerical procedure used in the present simulations are discussed. Then, the numerical solver is verified by simulating the decay of the Taylor-Green vortex, and high-order accuracy and low dissipation properties are confirmed in Sec.~\ref{sec:numerical_verification}. Results on the flow past the sphere at $Re$  = 10\,000 are discussed in Sec.~\ref{sec:LES_sphere}. The details of the simulation setup and the arrangement of cell types are described in Sec.~\ref{sec:model_setup}. A meticulous analysis was conducted of the instantaneous flow structures (Sec.~\ref{sec:flow_structures}) and mean flow topology (Sec.~\ref{sec:mean_flow_topology}), including the flow statistics around the sphere and in the near wake, and spectrum characteristics (Sec.~\ref{sec:spectrum_characteristics}).  Finally, Sec.~\ref{sec:conclusions} draws the conclusions of the paper.

\section{Numerical method}
\label{sec:numerical_model}
\subsection{Governing equations and solution procedure}
\label{sec:Governing_equations}
The LES approach is adopted in this work, for which the large-scale turbulence is resolved and a subgrid-scale model (SGS) is used to compute the effect of unresolved scales of turbulence. The governing equations are the spatially filtered incompressible Navier–Stokes equations, given as
\begin{equation}
	\nabla\cdot\mathbf{\widetilde{u}}=0, \label{eq-cont}
\end{equation}
\begin{equation}
	\frac{\partial\mathbf{\widetilde{u}}}{\partial t}+\nabla\cdot\left(\mathbf{\widetilde{u}}\otimes\mathbf{\widetilde{u}}\right)=-\nabla\widetilde{p}+\nabla\cdot\left[\nu\left(\nabla\mathbf{\widetilde{u}}+\nabla^{\mathrm{T}}\mathbf{\widetilde{u}}\right)\right]+\nabla\boldsymbol{\tau}^\mathrm{sgs},\label{eq-momnt1}
\end{equation}
where the tilde symbol $\widetilde{\cdot}$ denotes the spatial filtering over the grid in Cartesian coordinates $(x,y,z)$, $\mathbf{\widetilde{u}}=(\widetilde{u},\widetilde{v},\widetilde{w})$ is the filtered velocity vector, $\widetilde{p}$  is the filtered pressure divided by the density, and $\nu$ is the viscosity. In equation~\eqref{eq-momnt1}, the term $\boldsymbol{\tau}^{sgs}=(\widetilde{\boldsymbol{u}} \widetilde{\boldsymbol{u}}-\widetilde{\boldsymbol{u}\boldsymbol{u}})$ is the SGS stress tensor which can be expressed by an eddy viscosity model. Additionally, the deviatoric part is given by
\begin{equation}
	\boldsymbol{\tau}^{\mathrm{sgs}}-\frac{1}{3} \operatorname{trace}\left(\boldsymbol{\tau}^{\mathrm{sgs}}\right) \boldsymbol{I}=2 \nu_t \overline{\boldsymbol{S}}\label{eddy-viscosity}
\end{equation}
where $\overline{\boldsymbol{S}}=\frac{1}{2}\left(\nabla\mathbf{\widetilde{u}}+\nabla^{\mathrm{T}}\mathbf{\widetilde{u}}\right)$ is the strain rate tensor of the resolved field and $\boldsymbol{I}$ is the unit tensor.
For incompressible flows, the term $\operatorname{trace}\left(\boldsymbol{\tau}\right)$ is zero due to the divergence-free condition. 
By substituting \eqref{eddy-viscosity} into \eqref{eq-momnt1}, the momentum equation can be rewritten as follows
\begin{eqnarray}
	&  & \frac{\partial\mathbf{\widetilde{u}}}{\partial t}+\nabla\cdot\left(\mathbf{\widetilde{u}}\otimes\mathbf{\widetilde{u}}\right)=-\nabla\widetilde{p}+\nabla\cdot\left[\nu_{\mathrm{eff}}\left(\nabla\mathbf{\widetilde{u}}+\nabla^{\mathrm{T}}\mathbf{\widetilde{u}}\right)\right],\label{eq-momnt2}
\end{eqnarray}
where the total viscosity $\nu_{\mathrm{eff}}=\nu+\nu_t$ is the sum of the molecular viscosity and the eddy viscosity. The eddy viscosity $\nu_t$ is determined based on the energy content of the smallest resolved scale calculated with the dynamic Smagorinsky model (DSM) proposed by Lilly \cite{Lilly1991SGS}.

\begin{figure}[htbp]
	\centering
	\begin{overpic}[width=9cm]{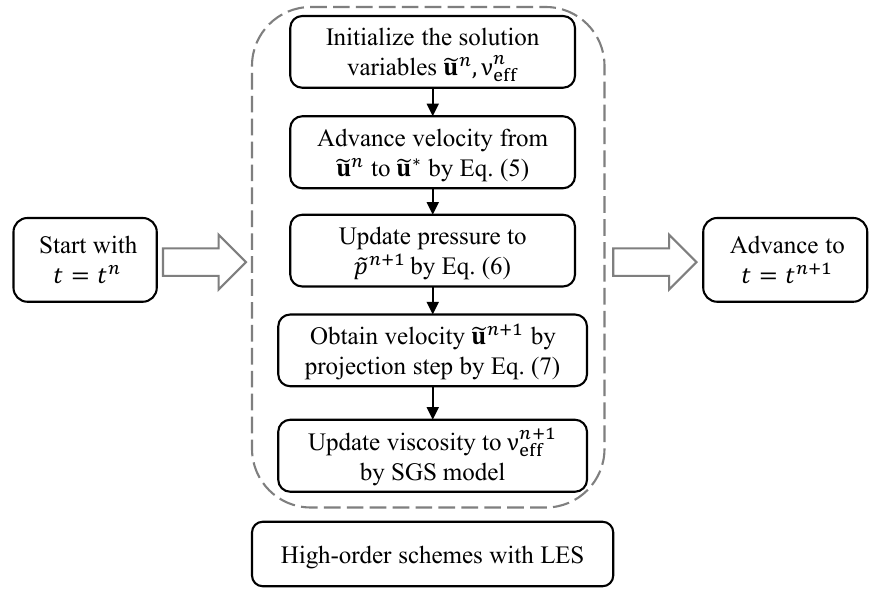}
	\end{overpic}		
	\caption[]{\textcolor{black}{A flow chart for the solution procedure of the present numerical model in a single time step.}}
	\label{fig:flow_chart_hybrid_method}
\end{figure}

Figure~\ref{fig:flow_chart_hybrid_method} gives a flow chart for the whole solution procedure of the numerical method in a time step. The solution procedure for the hybrid method is expressed in the semi-discrete formulations as follows. In the present framework, we use the fractional step approach \citep{Kim1985Application,Chorin1968Numerical,Harlow1965Numerical} to solve the governing equations of \eqref{eq-cont} and \eqref{eq-momnt2} with the third-order Total Variation Diminishing (TVD) Runge-Kutta (RK) methods \citep{Gottlieb1998TVD} that is used to advance the numerical solutions at the time level from $n$ $\ (t\ =\ t^n\ )$ to $n+1$ $\ (t\ =\ t^n\ + {\Delta t})$. Note that all the physical quantities default to the filtered value, and their tildes are omitted for brevity. The solution procedure of the present solver is expressed in the semi-discrete formulations as follows.

Step 1. Given the velocity field $\mathbf{u}^n$ at time level $t^n$, the intermediate velocity $\mathbf{u}^\ast$ is obtained by computing the convection and diffusion terms of the momentum equation,
\begin{eqnarray}
	&  & \frac{\partial\mathbf{u}}{\partial t}=-\nabla\cdot\left(\mathbf{u}\otimes\mathbf{u}\right)+\nabla\cdot\left[\nu_{\mathrm{eff}}\left(\nabla\mathbf{u}+\nabla^{\mathrm{T}}\mathbf{u}\right)\right].\label{eq-adv-diff}
\end{eqnarray}

Step 2. The pressure Poisson equation is constructed to compute the pressure field $p^{n+1}$ by enforcing the divergence-free condition,
\begin{equation}
	\nabla \cdot\left(\nabla p^{n+1}\right)=\frac{\nabla \cdot \mathbf{u}^{\ast}}{\Delta t}.\label{eq-poiss}
\end{equation}

Step 3. The velocity field is updated by the projection step to satisfy the continuity equation,
\begin{equation}
	\mathbf{u}^{n+1}=\mathbf{u}^{\ast}-\nabla p^{n+1}\Delta t.\label{eq-proj}
\end{equation}
The velocity $\mathbf{u}^{n+1}$ satisfying the divergence condition is utilized to calculate the transport velocity in the subsequent time iteration.

\subsection{The computational grid and variable definition}
\begin{figure}[!htpb]
	\begin{minipage}[c]{1\linewidth}
		\begin{overpic}[height=5cm]{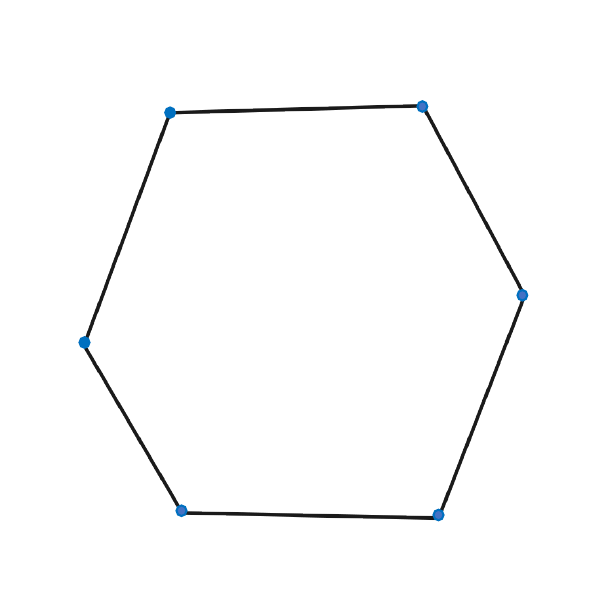}
			\put(0,80){\color{black}{(a)}}
            \put(48,43){\color{black}{$\Omega_{i}$}}
            \put(68,85){\color{black}{$\theta_{i1}$}}
            \put(26,84){\color{black}{$\theta_{i6}$}}
            \put(90,49){\color{black}{$\theta_{i2}$}}
            \put(6,40){\color{black}{$\theta_{i5}$}}
            \put(75,8){\color{black}{$\theta_{i3}$}}
            \put(26,8){\color{black}{$\theta_{i4}$}}
            \put(82,28){\color{black}{$\Gamma_{ij}$}}
		\end{overpic}
        \begin{overpic}[height=5cm]{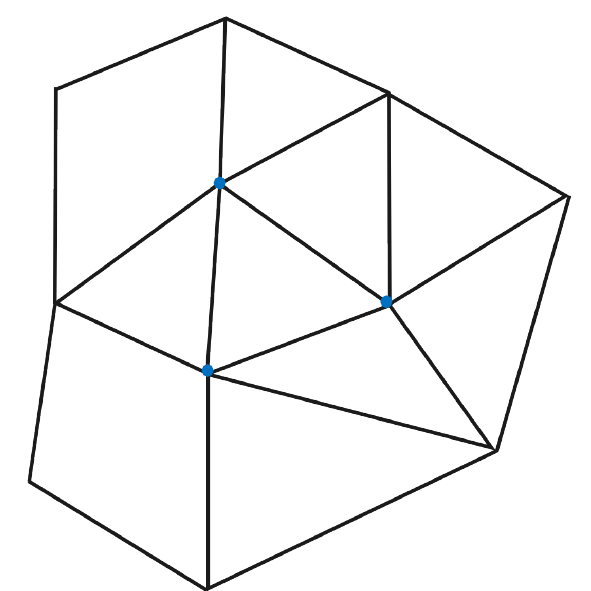}
            \put(0,90){\color{black}{(b)}}
            \put(46,50){\color{black}{$\Omega_{i}$}}
            \put(60,36){\color{black}{$\Omega_{ikl}$}}
            \put(68,46){\color{black}{$\theta_{ik}$}}
        \end{overpic}
	\end{minipage}
	\caption[]{(a) Sketch of the control volume and (b) reconstruction template for two-dimensional unstructured meshes. $\Omega_{i}$ denotes the target cell and  $\Omega_{ikl}$ represents its neighboring cells which share the common vertex $\theta_{ik}$.}
	\label{fig:grid_element}
\end{figure}
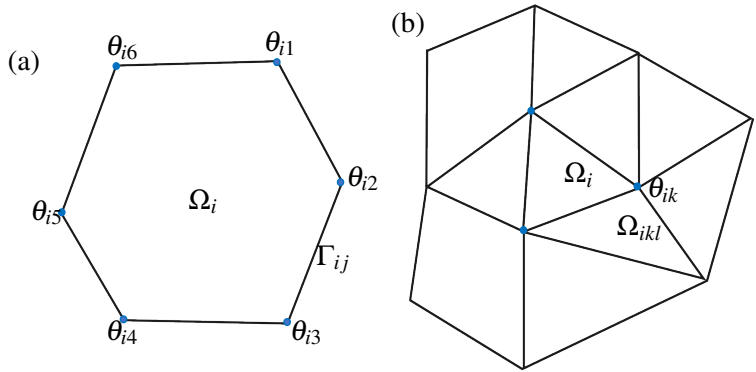

We use the proposed high-order finite volume method FVMS3 assigned to polygonal or polyhedral meshes with an arbitrary number of faces and vertices. As shown in Fig.~\ref{fig:grid_element}(a), the computational domain is divided into a set of non-overlapping control volumes ${\Omega}_i\ (i\ =\ 1,2\ldots,I)$. For each grid element ${\Omega}_i$, the connectivity information is crucial to represent the mesh, which contains: a list of vertices $\theta_{ik}(k\ =\ 1,2,...,K)$ and the boundary edges/surfaces ${\Gamma}_{ij}\ (j\ =\ 1,2,...,J)$ with their unit normal vector ${\mathbf{n}}_{ij}$. For the topological relation of a given element ${\Omega}_i$, ${\Omega}_{ikl}$ and ${\Omega}_{ij}$ denote the neighboring elements sharing the common vertex $\theta_{ik}$ and the one sharing the boundary surface ${\Gamma}_{ij}$, respectively. For later use, we denote the coordinate of the mass center $\theta_{ic}$ by $(x_{ic},y_{ic},z_{ic})$, the size of the boundary surface by $\left|\Gamma_{ij}\right|$, and the mass volume by $\left|\Omega_{i}\right|$. 

In a collocated finite volume method, the numerical variables in a cell ${\Omega}_i$ are defined by the Volume Integrated Average (VIA) as
\begin{equation}
	\label{eq:VIA}
	\bar{\phi}_{i}(t) \equiv \frac{1}{\left|\Omega_{i}\right|} \int_{\Omega_{i}} \phi(x, y, z, t) d \Omega,
\end{equation}
where $\phi(x,y,z,t)$ denotes any physical quantity of interest, such as velocity and pressure. It should be noted that Gaussian quadrature points are employed in order to perform numerical integration of the quantity $\phi$ over both the cell volume and the boundary surface.

\subsection{High-order reconstruction algorithm}
Given VIA, the reconstruction of the discrete solution variable can be achieved through a piecewise quadratic polynomial in three dimensions of the general form:  
\begin{equation}\small
	\Phi_{i}(\phi|; \mathbf{x})  =  \stackrel{(\alpha+\beta+\gamma)\leqslant 2}{\sum_{\alpha=0}\sum_{\beta=0}\sum_{\gamma=0}}(x-x_{ic})^{\alpha}(y-y_{ic})^{\beta}(z-z_{ic})^{\gamma}c_{\alpha\beta\gamma},\label{eq:scheme_poly}
\end{equation}
where $c_{\alpha\beta\gamma}(0\leqslant\alpha,\beta,\gamma\leqslant 2)$
denotes the undetermined coefficients of a quadratic polynomial. For the high-order method on the unstructured grid, a large template is constructed from allowed cells to contain sufficient degrees of freedom (DOFs), which should be given special attention to overcome the difficulties associated with algorithmic complexity and computational efficiency.

As shown in Fig.~\ref{fig:grid_element}(b), the FVMS3 scheme builds a compact reconstruction template by merging all of the neighboring cells that share the common vertices $\theta_{ik}$, which can be described by
\begin{equation}
	\left\{\Omega_{is} \mid: s=0,1, \dot, S\right\} \equiv \stackrel{k=1}{K} \cup \stackrel{l=1}{L} \cup \Omega_{ikl}.\label{stencil}
\end{equation}
The allowed cells in the supporting stencil are re-labeled with a new set of indices, designated "$is$", which range from one to the total number of cells. The subscript $\Omega_{i0}$ denotes the target cell. A connection table is employed to store the topological relations between the local and global indices for each stencil.

\subsection{The spatial discretization using the finite volume method}
\color{black}
For the spatial discretization, we first solve Eq.~\eqref{eq-adv-diff} through a finite volume scheme that converts the numerical fluxes of the convection and diffusion terms by summing them across each boundary segment,
\begin{equation}
\footnotesize
	\frac{\partial \overline{\mathbf{u}}_i}{\partial t}=\frac{1}{\left|\Omega_i\right|}\left(-\sum_{j=1}^J\left(v_{n_{ij}} \mathbf{u}_{ij}+(\nu_{\mathrm{eff}})_{ij}\left((\nabla \mathbf{u})_{ij}+(\nabla^{\mathrm{T}}\mathbf{u})_{ij}\right) \cdot \mathbf{n}_{i j}\right)\left|\Gamma_{ij}\right|\right)
\end{equation}
where $v_{n_{ij}}=\hat{\mathbf{u}}_{ij}^n \cdot \mathbf{n}_{ij}$ is the transport velocity. 
The surface integrated average values $\mathbf{u}_{ij}$ are computed from the reconstruction functions in the two adjacent cells using an upwind scheme,
\begin{equation}
\mathbf{u}_{ij} = \frac{1}{|\Gamma_{ij}|} \int_{\Gamma_{ij}} \Phi_{iup}(\mathbf{u}|; \mathbf{x}) d\Gamma_{ij},
\label{eq:advection}
\end{equation}
where $\Phi_{iup}$ denotes the reconstruction function over the upwinding cell, i.e.
\begin{equation}
iup = 
\begin{cases} 
i, & \text{for } \mathbf{u}_{ij} \cdot \mathrm{n}_{ij} \geq 0; \\
j, & \text{otherwise}.
\end{cases}
\end{equation}
The surface integrated average gradients $(\nabla \mathbf{u})_{ij}$ are obtained from the reconstruction polynomial of Eq.~\eqref{eq:scheme_poly} which are approximated at the boundary surface  by
\begin{equation}
\small
\begin{aligned}
(\nabla\mathbf{u})_{ij} =& \frac{1}{2} (\nabla\mathbf{u}^+_{ij} + \nabla\mathbf{u}^-_{ij}) \\
&+ \frac{\mathbf{r}_{ij}}{|\mathbf{r}_{ij}|} \left( \frac{(\mathbf{u}_{jc} - \mathbf{u}_{ic})}{|\mathbf{r}_{ij}|} - \frac{1}{2} (\nabla\mathbf{u}^+_{ij} + \nabla\mathbf{u}^-_{ij}) \cdot \mathbf{r}_{ij} \right),
\label{eq:gradient_BC}
\end{aligned}
\end{equation}
where $(\nabla\mathbf{u})^{\pm}_{ij}$ denotes the velocity gradients reconstructed at the boundary surface $\Gamma_{ij}$ in the target cell $\Omega_i$ by
\begin{equation}
(\nabla\mathbf{u})^+_{ij} = \frac{1}{|\Gamma_{ij}|} \int_{\Gamma_{ij}} \nabla \Phi_i (\mathbf{u}|;\mathbf{x}) d\Gamma_{ij},
\end{equation}
and in the adjacent cell $\Omega_j$ by
\begin{equation}
(\nabla\mathbf{u})^-_{ij} = \frac{1}{|\Gamma_{ij}|} \int_{\Gamma_{ij}} \nabla \Phi_j (\mathbf{u}|;\mathbf{x}) d\Gamma_{ij},
\end{equation}
respectively. The extra correction term on the right-hand side of Eq.~\eqref{eq:gradient_BC} employs reconstructed values, denoted as $\mathbf{u}_{i(j)c}$, at the mass center. This treatment has been proven to greatly improve the numerical accuracy and dissipation properties over conventional second-order schemes. The pressure field is then solved from the Poisson equation, with the pressure gradient approximated by linear interpolation using a semi-implicit formulation. After obtaining the pressure field $p^{n+1}$, the velocity is updated using the time-evolution-converting (TEC) formula. For more algorithm details, refer to Xie \textit{et al.} \cite{Xie2019High-fidelitysolver,Xie2020consistent}. 

\color{black}
\section{Numerical verification: Taylor-Green vortex problem}
\label{sec:numerical_verification}
\subsection{Numerical accuracy and convergence behavior}
The high-order numerical solver is verified by the decay of the Taylor–Green vortex \citep{Ethier2007taylor,Ferrer2011taylor} for incompressible Navier-Stokes equations in which analytical solutions are given by
\begin{gather}
	u(x,y,t)  =  -\cos(\pi x)\sin(\pi y)\exp\left(-\frac{2\pi^{2}t}{\operatorname{Re}}\right),\\[3pt]
	v(x,y,t)  =  \sin(\pi x)\cos(\pi y)\exp\left(-\frac{2\pi^{2}t}{\operatorname{Re}}\right),\\[3pt]
	p(x,y,t)  =  -\frac{1}{4}(\cos(2\pi x)+\sin(2\pi y))\exp\left(-\frac{2\pi^{2}t}{\operatorname{Re}}\right).
\end{gather}
The simulation is carried out on a square domain of $[-1,1]^2$ divided into Delaunay triangular elements with doubled resolution. The Reynolds number, $Re=100$, was initially employed to compute the solution, with a uniform time step of $\Delta t=1\times10^{-4}$ s. 
\textcolor{black}{
To quantify the solution accuracy of the numerical results, we define the errors of $L_2$ norms as follows:
\begin{equation}
\begin{aligned}
E\left(L_2\right) & =\sqrt{\frac{\sum_{i=1}^N\left(\left|\Phi_{\mathrm{n} i}-\Phi_{\mathrm{e} i}\right|^2\left|\Omega_i\right|\right)}{\sum_{i=1}^N\left|\Phi_{\mathrm{e} i}\right|^2\left|\Omega_i\right|}},
\end{aligned}
\end{equation}
where $N$ is the total number of mesh cells, $\Phi_{\mathrm{n} i}$ and $\Phi_{\mathrm{e} i}$ represent the numerical and exact solutions, respectively.}

\begin{figure}[!htpb]
	\centering
	\begin{minipage}[c]{1\linewidth}
		\centering
		\begin{overpic}[width=6cm]{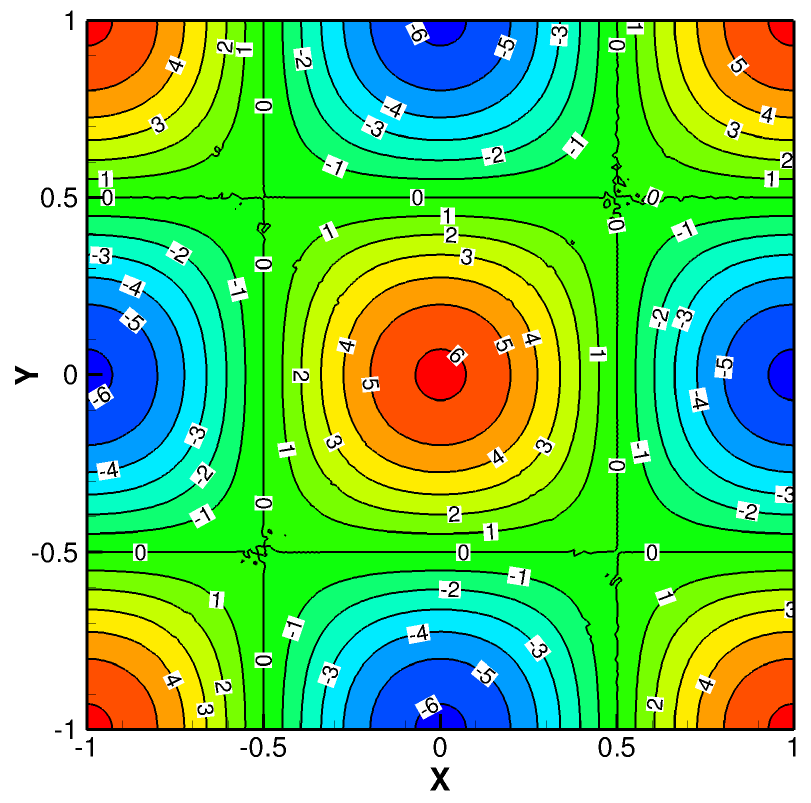}
			\put(-2,94){\color{black}{(a)}}
		\end{overpic}
		\qquad
		\begin{overpic}[width=6cm]{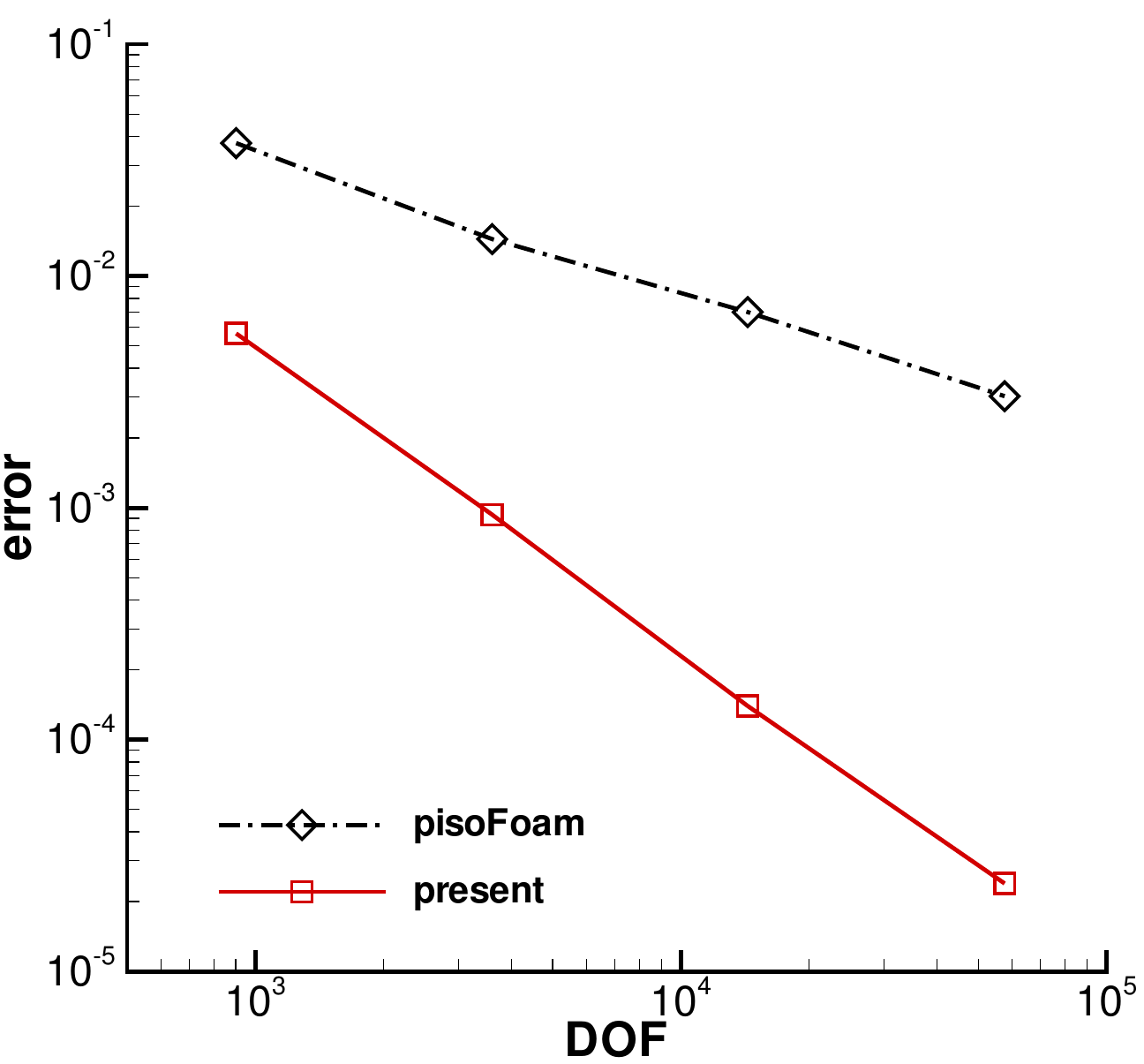}
			\put(-4,92){\color{black}{(b)}}
		\end{overpic}
	\end{minipage}
	\caption[]{Results obtained at time $t=0.1$~s by the present solver of the Taylor-Green vortex decay. (a) The contour of the vorticity and (b) the $L_{2}$ errors of the velocity against the DOFs together with those computed by the pisoFoam solver.}
	\label{fig:taylor_vortex}
\end{figure}
The contour of the vortex at time $t=0.1$~s is shown in Fig.~\ref{fig:taylor_vortex}(a), which is accurately predicted by the present solver on a mesh of $57\ 518$ cells. We also present the $L_{2}$ errors of velocity versus DOFs in Fig.~\ref{fig:taylor_vortex}(b), compared with results from pisoFoam with the conventional QUICK (Quadratic Upwind Interpolation for Convective Kinematics) scheme \citep{Shterev2012quick}. The results suggest that the $L_{2}$ errors of the present numerical method are reduced to almost three orders of magnitude, which shows considerable improvements both in numerical accuracy and convergence behavior. The velocity convergence rate is found to be higher than 2.5. The present solver is capable of providing high-order solutions for the accurate simulation of incompressible flows, thereby justifying its utility.

\begin{figure}[!htpb]
	\centering
	\begin{minipage}[c]{1\linewidth}
		\centering
		\begin{overpic}[width=6cm]{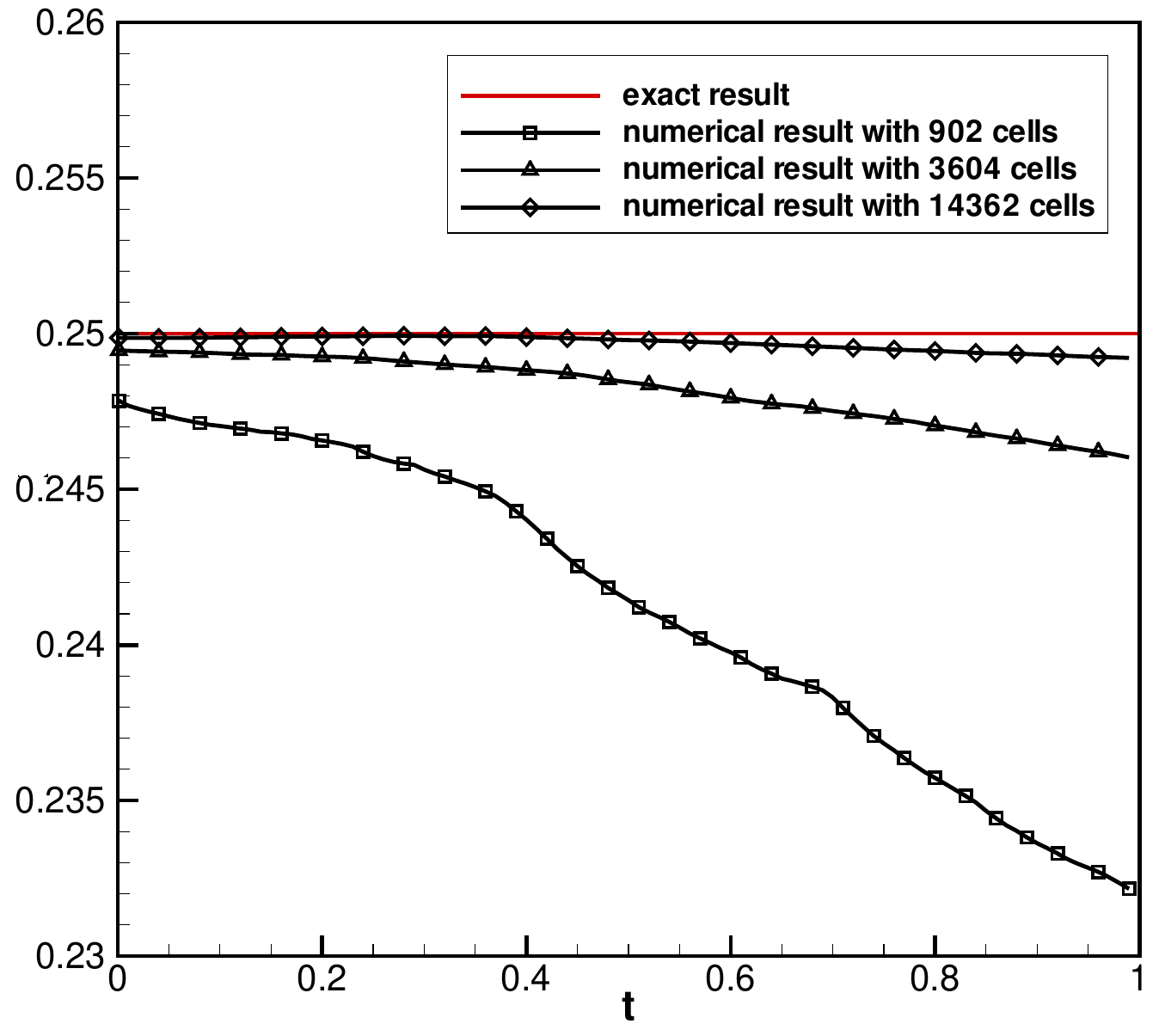}
			\put(-6,86){\color{black}{(a)}}
		\end{overpic}
		\qquad
		\begin{overpic}[width=6cm]{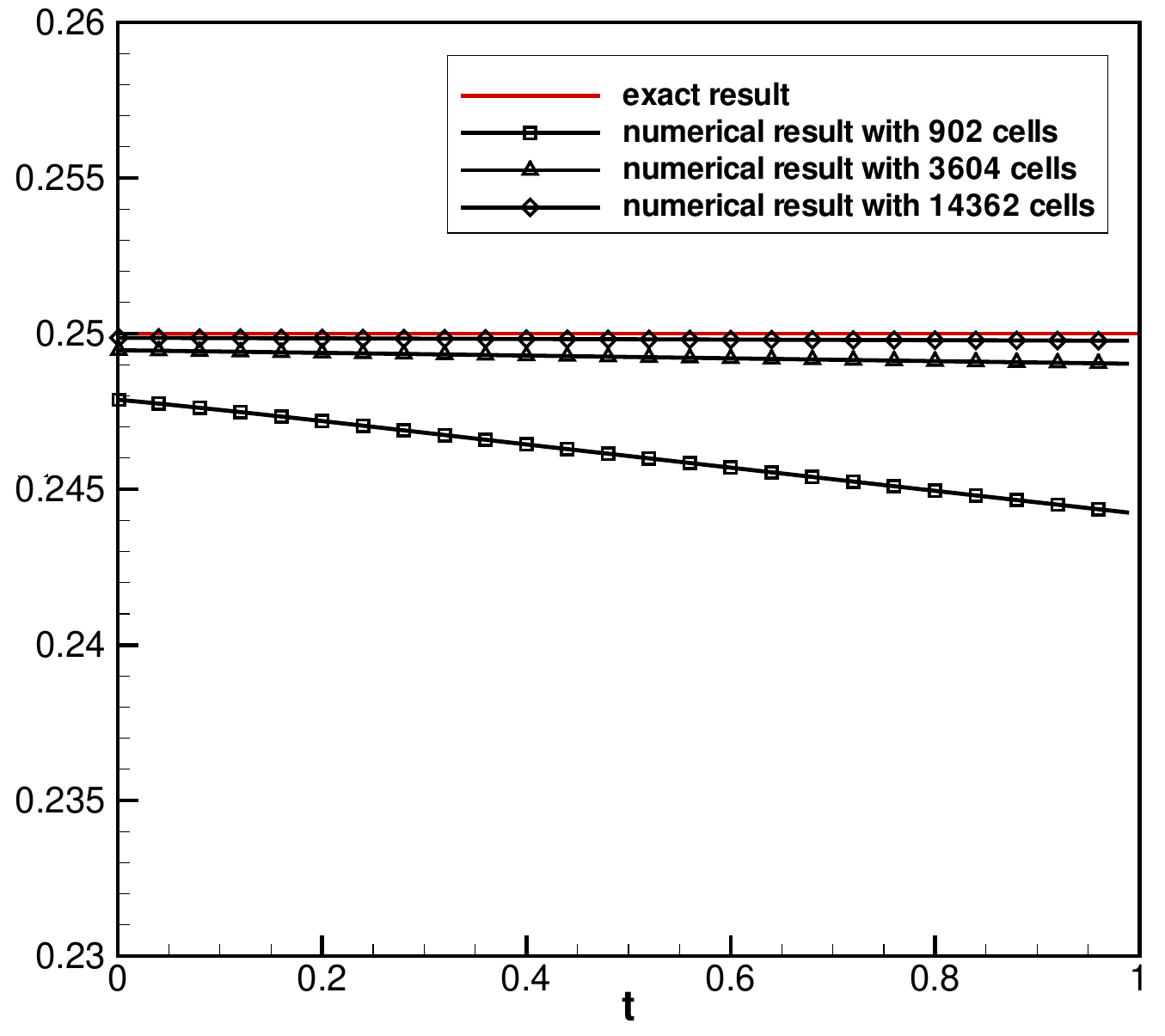}
			\put(-6,86){\color{black}{(b)}}
		\end{overpic}
	\end{minipage}
	\caption[]{The evolution of the total kinetic energy $K_{\text{num}}(t)$ for the inviscid TGV computed by (a) pisoFoam \citep{xie2017piso} and (b) the present solver on the different grid resolutions.}
	\label{fig:kinetic_scheme}
\end{figure}
\subsection{ Energy conservation and numerical dissipation}
For high-Reynolds number flows, the conservation of kinetic energy is of critical importance for the accurate simulation of flow fields. As the numerical dissipation may potentially overwhelm the impact of the turbulent viscosity model within the context of LES, further investigation into the relative effectiveness of the energy conservation property is warranted, as reported by Perot\cite{Perot2011conservation,HAO2024LES}. To assess the kinetic energy change, it is necessary to define the total kinetic energy within the computational domain, which is dependent upon the Reynolds number and varies over time.
\begin{equation}\small
	K_{\text {ext}}(t)=\frac{1}{2|\Omega|} \int_{-1}^1 \int_{-1}^1\left(u^2+v^2\right)dx dy=\frac{1}{4} \exp \left(-\frac{4 \pi^2 t}{\operatorname{Re}}\right).
\end{equation}
In the limit of inviscid incompressible flow, the kinetic energy of the TGV should be exactly conserved as a constant value of 1/4. We perform the simulation on the different grid resolutions up to time $t=1$~s. To measure the numerical dissipation, we define the total kinetic energy $K_{\text{num}}$ and the numerical dissipation rate $\varepsilon$ of the numerical solutions as 
\begin{gather}
	K_{\text{num}}(t)=\frac{1}{\sum\left|\Omega_i\right|}\sum_i\left(\frac{\bar{u}_i(t)^2+\bar{v}_i(t)^2}{2}\left|\Omega_i\right|\right),\\[3pt]
	\varepsilon=\frac{K_{\text{num}}(t=0)-K_{\text{num}}(t=1)}{K_{\text{num}}(t=0)},
\end{gather}

\begin{table*}[!t]\small
    \caption{The kinetic energy $K_{\text{num}}(t)$ and the dissipation rate $\varepsilon$ of the numerical solutions computed by pisoFoam \citep{xie2017piso} and the present solver on the different grid resolutions.}
    \label{tab:kenetic_energy}
    \begin{center}
        \def~{\hphantom{0}}
        \begin{ruledtabular}
            \begin{tabular}{lccccc}
                \multirow{2}{*}{Elements} & \multirow{2}{*}{\textcolor{black}{Ref. solutions}} & \multicolumn{2}{c}{pisoFoam} & \multicolumn{2}{c}{Present solver} \\ \cline{3-4} \cline{5-6}
                & & $K_{\mathrm{num}}$ ($t=1$~s) & {\textcolor{black}{Relative Error $\varepsilon$}} & $K_{\mathrm{num}}$ ($t=1$~s) & {\textcolor{black}{Relative Error $\varepsilon$}} \\
                \hline
                902 & 0.247876 & 0.232162 & \textcolor{black}{$6.35\%$}  & 0.244248 & \textcolor{black}{$1.47\%$}   \\
                3604 & 0.249466 & 0.246024 & \textcolor{black}{$1.38\%$} & 0.249034 & \textcolor{black}{$0.17\%$}   \\
                14\,362 & 0.249867 & 0.249221 & \textcolor{black}{$0.26\%$} & 0.249772 & \textcolor{black}{$0.04\%$}   \\
            \end{tabular}
        \end{ruledtabular}
    \end{center}
\end{table*}
where $\left|\Omega_i\right|$ is the volume of each element. As shown in Fig.~\ref{fig:kinetic_scheme}, the present solver shows significant improvements in preserving the conservation of total kinetic energy, which has much less numerical dissipation than pisoFoam and converges rapidly with mesh refinement.
\textcolor{black}{More quantitatively, we also present the kinetic energy $K_{\text{num}}$ and its relative error at $t=0$ s in Table~\ref{tab:kenetic_energy}, which further serves as an indicator of the numerical dissipation rate $\varepsilon$ in the numerical solutions.} Compared to pisoFoam, the kinetic energy of the present solver is adequately conserved with nearly 1/10 dissipation rate which can be visually confirmed in Fig.~\ref{fig:kinetic_scheme}. By the high-order accuracy and low-dissipation property of the FVMS3 scheme, the present solver is considered to be one of the practical and attractive choices for accurate simulation of incompressible turbulent flows around the sphere, which will be described in the subsequent section.

\subsection{Comparison with high-order codes}
\label{sec:high-order codes}
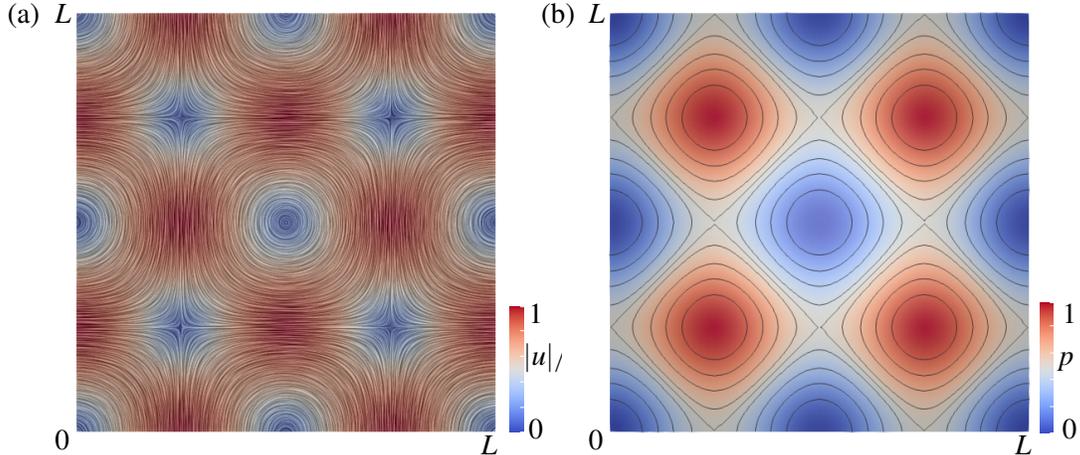
\begin{figure}[!htpb]
	\centering
	\begin{minipage}[c]{1\linewidth}
		\centering
		\begin{overpic}[trim=1cm 1cm 1cm 1cm, clip, width=7cm]{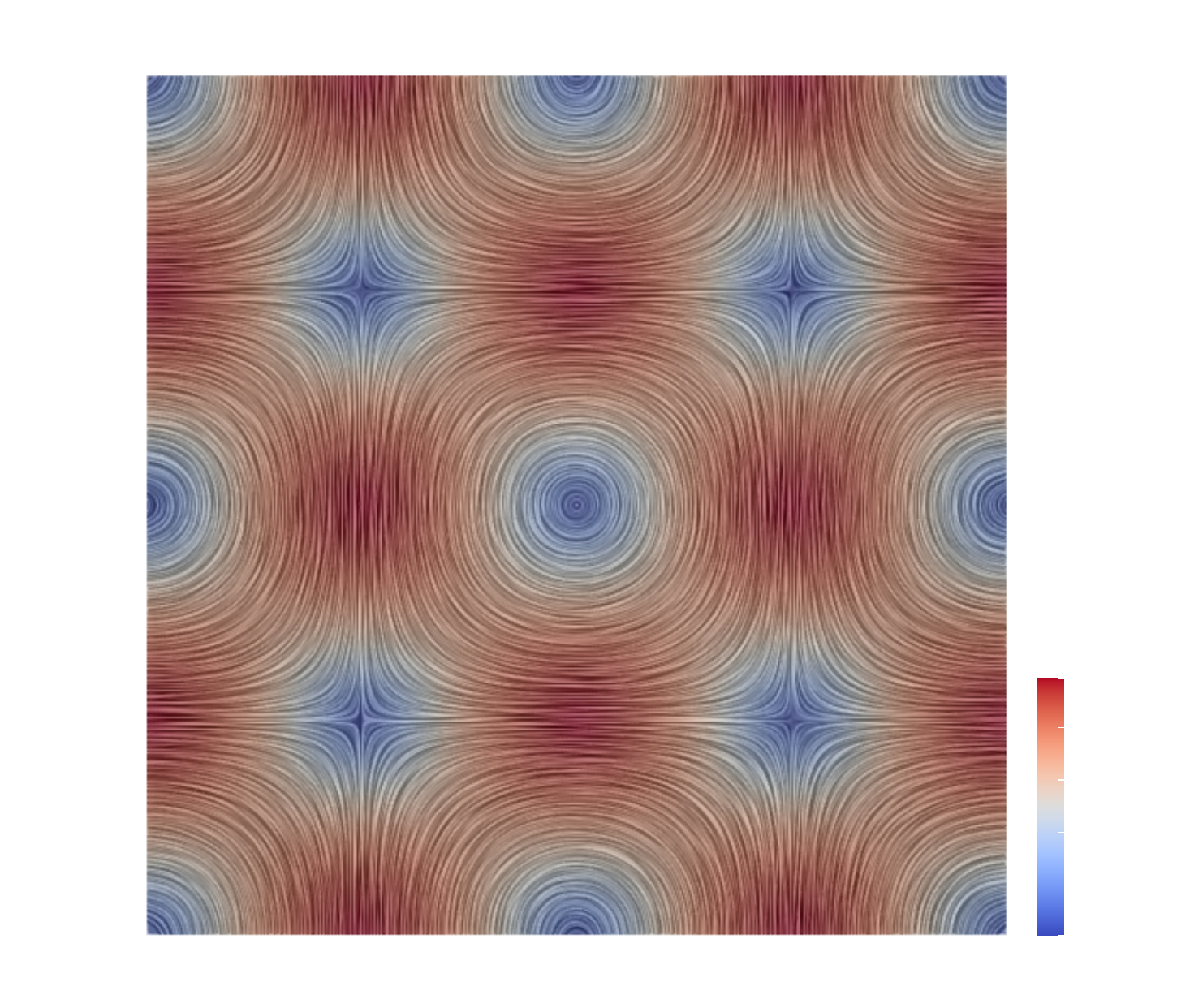}
			\put(-4,80){\color{black}{(a)}}
            \put(86,-2){\color{black}{$L$}}
            \put(5,-1){\color{black}{0}}
            \put(5,80){\color{black}{$L$}}
            \put(94,15){\color{black}{$|u|/u_0$}}
            \put(95,1){\color{black}{0}}
            \put(95,23){\color{black}{1}}
		\end{overpic}
		\begin{overpic}[trim=1cm 1cm 1cm 1cm, clip, width=7cm]{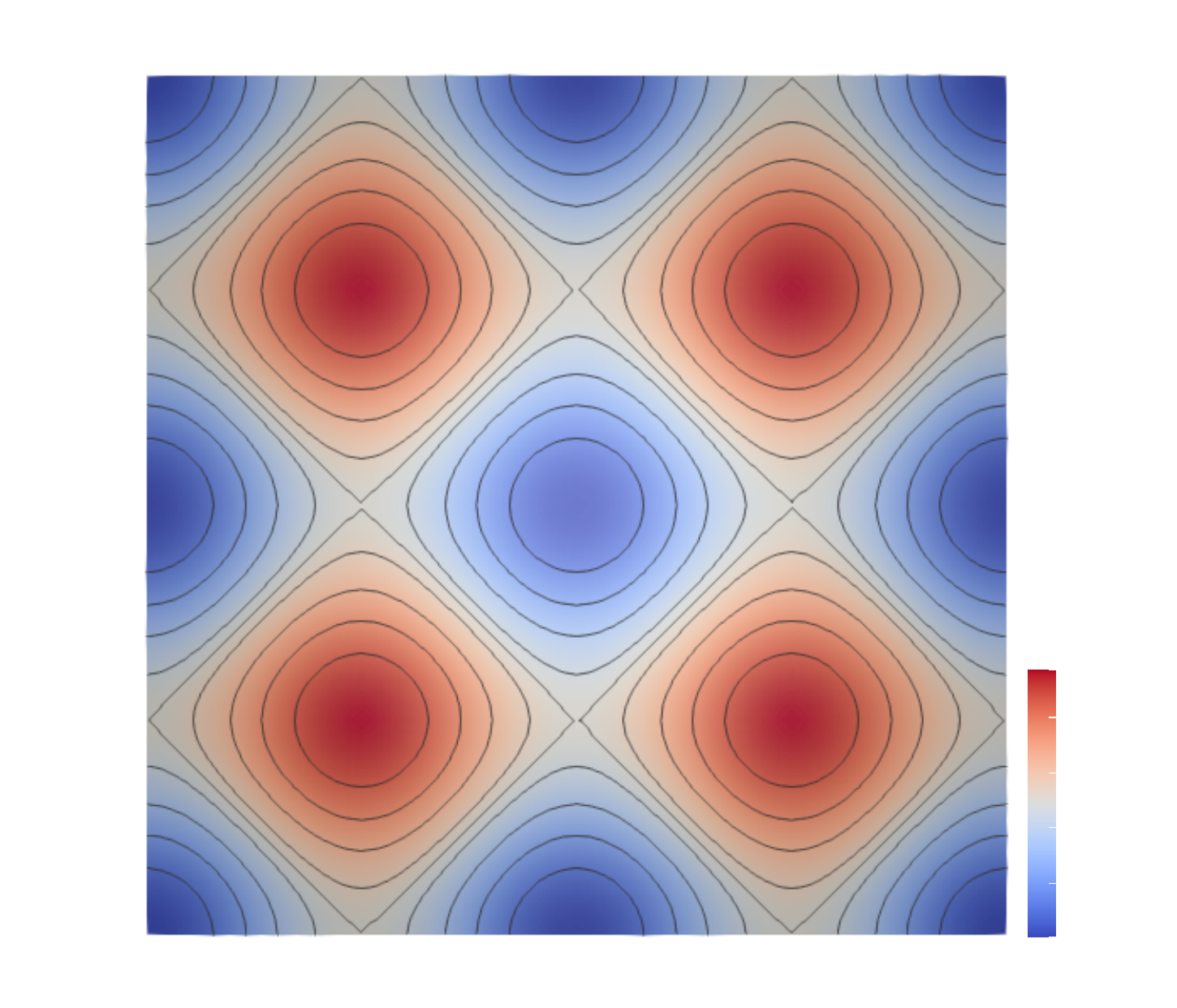}
			\put(-4,80){\color{black}{(b)}}
            \put(86,-2){\color{black}{$L$}}
            \put(5,-1){\color{black}{0}}
            \put(5,80){\color{black}{$L$}}
            \put(94,15){\color{black}{$p$}}
            \put(95,1){\color{black}{0}}
            \put(95,23){\color{black}{1}}
		\end{overpic}
	\end{minipage}
	\caption[]{\textcolor{black}{Results of the 2D TGV at $Re$ = 1\,600 simulated by the present FVMS3 solver at time $t=10$ including (a) velocity LIC colored by the non-dimensioned velocity magnitude, and (b) the corresponding pressure contour.}}
	\label{fig:TGV_vortex_Re1600}
\end{figure}

\begin{table*}[!htpb]\small
  \caption{\textcolor{black}{Comparison of maximum velocities and relative errors for different solvers. Note that the numerical results for DINO and Nek5000 are extracted from the work of Abdelsamie \textit{et al.} (2021). \cite{abdelsamie2021taylor} }}
  \label{tab:Solver_comparison}
    \centering
    \begin{ruledtabular}
      \textcolor{black}{
        \begin{tabular}{lccccc}
          & Analytical & pisoFoam & Present: FVMS3 & DINO & Nek5000 \\ \hline
          $V_{\text{max}}$  &  0.987578 & 0.978258 & 0.986686 & 0.987565 & 0.987578 \\
          $\varepsilon_{\text{rel}}$ & (Ref) & $9.4\times 10^{-1} \%$ & $9.0\times 10^{-2} \%$ &  $1.3 \times 10^{-3} \%$ & $7.1 \times 10^{-5} \%$ \\
          Discretization Type & $\backslash$ & FVM & FVM & FDM & SEM \\
          Spatial order & $\backslash$  & 2nd & 3rd & 6th & 8th \\ 
          Temporal scheme & $\backslash$  & backward & impl. RK3 & expl. RK4 & semi-impl. BDF3\\
          Grid type & $\backslash$ & Polyhedral & Polyhedral & Hexahedral & Hexahedral\\
        \end{tabular}
      }
    \end{ruledtabular}
\end{table*}
\textcolor{black}{While validations have been established using a low-order numerical solver, this section expands the comparison to include solvers of various orders, demonstrating the robustness and superiority of the proposed method across a broader range of alternatives. Comparisons are made between the present solver and the second-order pisoFoam, as well as three high-order codes: DINO and Nek5000 \cite{nek5000_v17}. The analytical solution serves as the reference. Given that these codes have been extensively discussed in the literature and are not entirely new, only the most pertinent features are presented here. The investigated problem is the 2D Taylor-Green Vortex (TGV) at $Re=1\,600$. To facilitate direct comparison with previous results, all physical conditions and numerical setups are identical to those used by Abdelsamie \textit{et al.} (2021) \cite{abdelsamie2021taylor}. Simulations are conducted in a square domain with dimensions $[0, L]^2$ and periodic boundary conditions in both directions. Here, $L = 2\pi L_0$ with $L_0 = 1$m. The time step is $\Delta t = 5 \times 10^{-4} \text{s}$, and $N = 64$ grid points are used in each direction. The flow is simulated for a physical time of $t = 10$s.} 

\textcolor{black}{Figure~\ref{fig:TGV_vortex_Re1600} presents the flow topologies and pressure contour at $t = 10 \, \text{s}$. The vortex structures are visualized using the Line Integral Convolution (LIC) technique \cite{Loring2015Numerical}, which convolves noise with a vector field to produce streak patterns following the vector field tangents. Table~\ref{tab:Solver_comparison} compares maximum velocities and relative errors for different solvers, utilizing various orders and discretization methods, including the Finite Volume Method (FVM), Finite Difference Method (FDM), and Spectral Element Method (SEM). The results show that the present third-order FVM solver performs better than the second-order pisoFoam. The two other solvers demonstrated greater accuracy for this problem since higher-order schemes are used for spatial discretization. It is noted that these solvers require regular hexahedral meshes and their potential applications are limited. It is encouraging to note that the present scheme delivers high-order numerical accuracy regardless of grid types which is appealing for practical application with complex geometries.}

\subsection{Numerical dissipation on different cell types}
\begin{figure*}[!t]
  \centering
  \begin{minipage}[c]{1\linewidth}
  \centering
  \begin{overpic}[trim=3cm 3cm 3cm 3cm, clip, width=4.5cm]{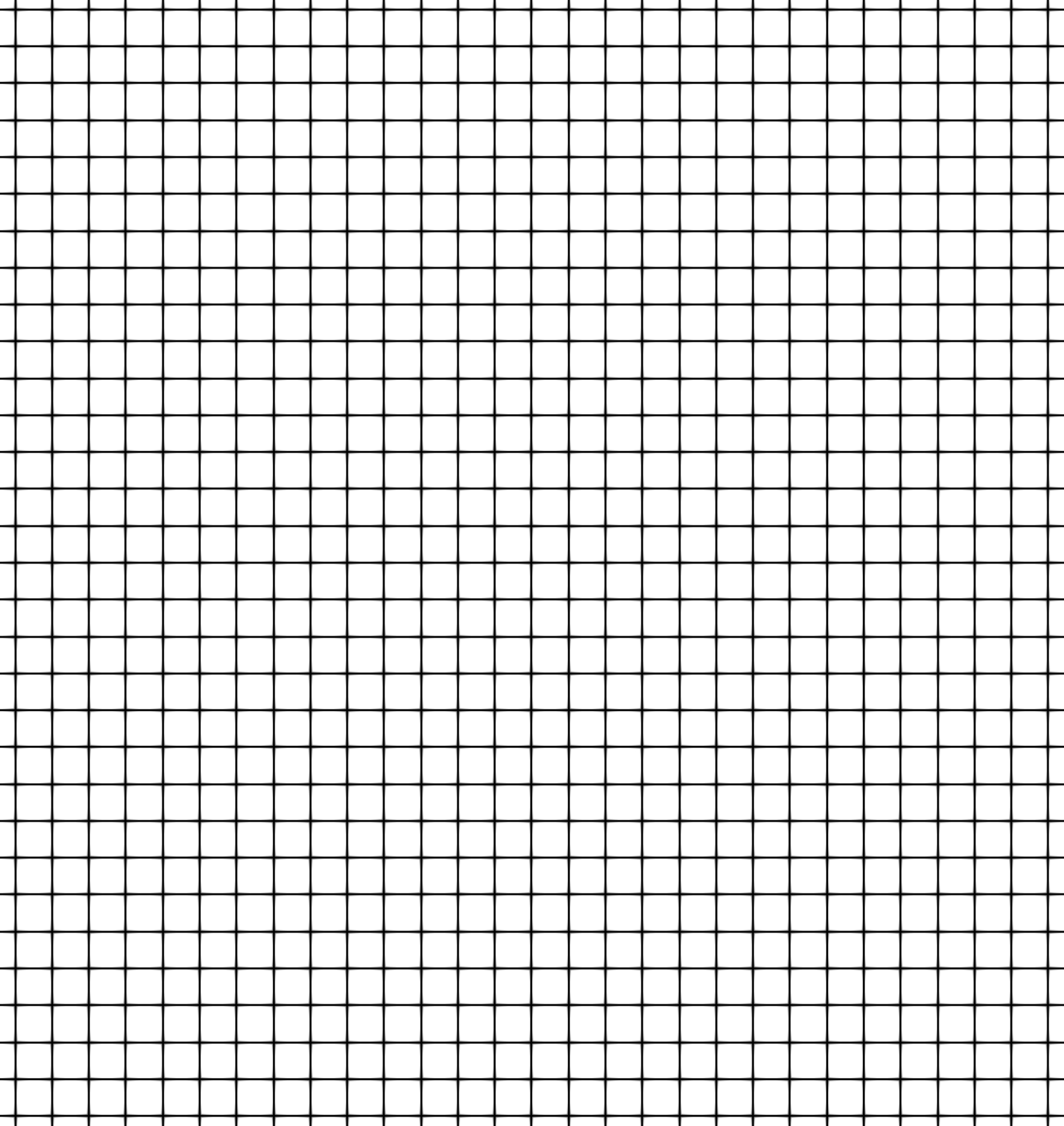}
  \put(-10,95){\color{black}{(a)}}
  \end{overpic}
\qquad
  \begin{overpic}[trim=3cm 3cm 3cm 3cm, clip, width=4.5cm]{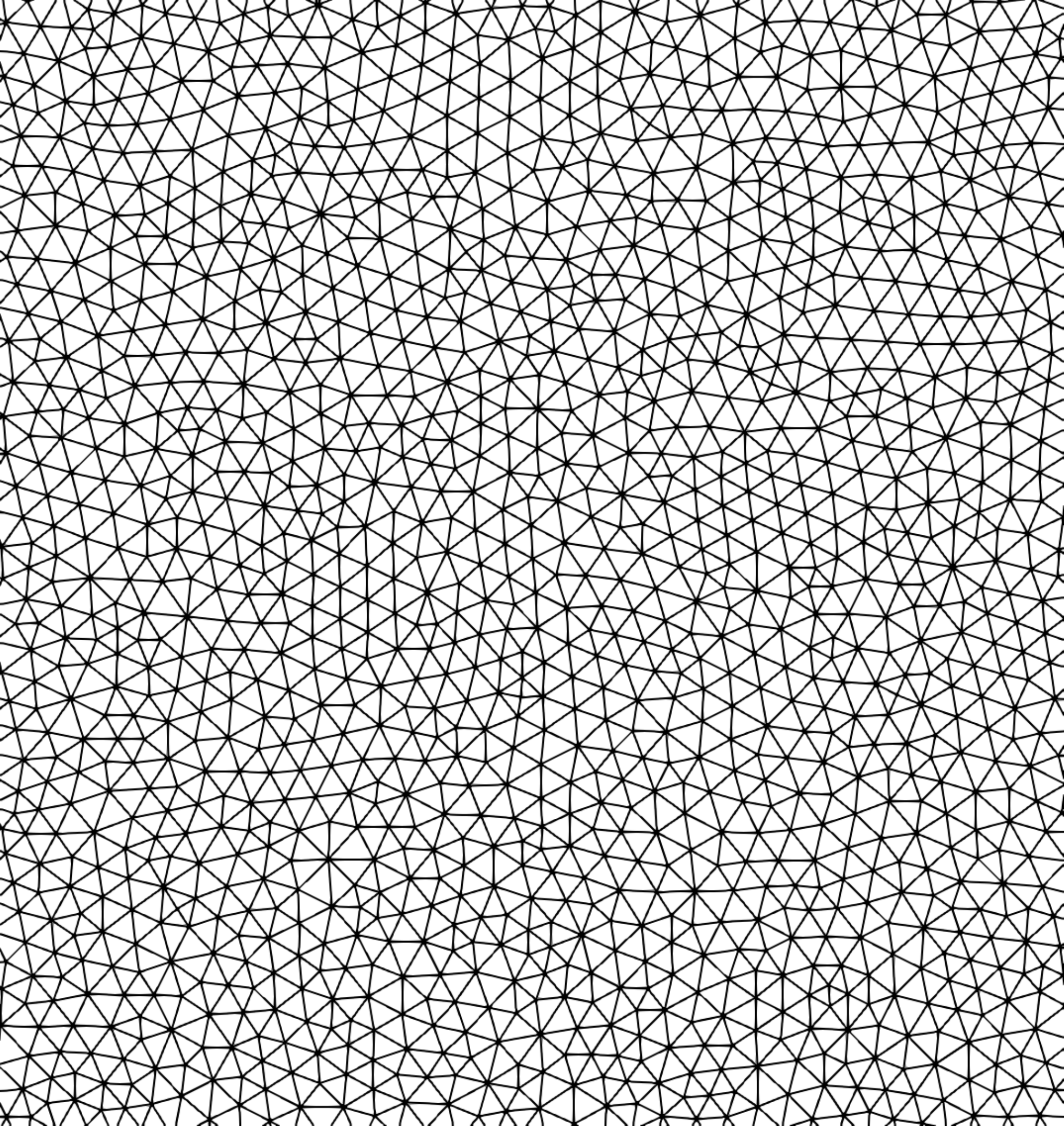}
  \put(-10,95){\color{black}{(b)}}
  \end{overpic}
\qquad
  \begin{overpic}[trim=3cm 3cm 3cm 3cm, clip, width=4.5cm]{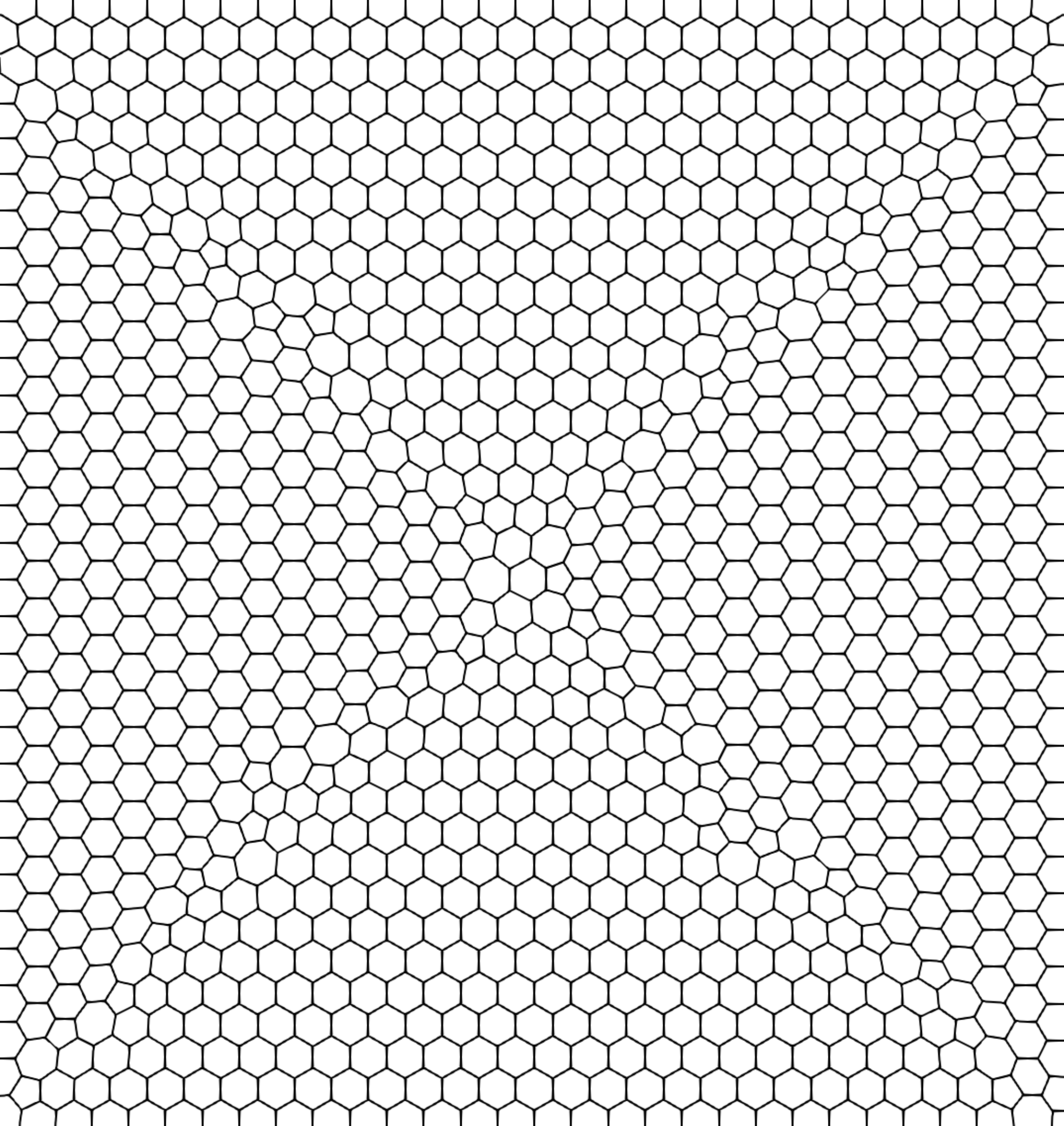}
  \put(-10,95){\color{black}{(c)}}
  \end{overpic}
  \end{minipage}
	\caption[]{\textcolor{black}{Computation grids used for benchmark tests. (a) Cartesian quadrilateral grid (grid-I), (b) Delaunay triangular grid (grid-II), and (c) polyhedra grid (grid-III).}}
	\label{fig:cell_types}
\end{figure*}

\begin{table*}[!htpb]\small
  \caption{\textcolor{black}{Kinetic energy and numerical dissipation of Taylor vortex problem at $R$e=1\,600 computed by pisoFoam and FVMS3 schemes on different grids for a physical time of $t = 10$ s.}}
  \label{tab:Solver_comparison_on_grids}
    \centering
    \begin{ruledtabular}
      \textcolor{black}{
        \begin{tabular}{lcccccccc}
        \multirow{2}{*}{} & \multirow{2}{*}{Cells} & Analytical   & \multicolumn{3}{c}{pisoFoam}     & \multicolumn{3}{c}{Present: FVMS3}        \\ \cline{4-6}  \cline{7-9}
                          &                       & $K_\mathrm{ref.}$  & $K_\mathrm{num}$  & Dissipation ($\varepsilon$) & Execution time (s)    & $K_\mathrm{num}$  & Dissipation ($\varepsilon$) & Execution time (s)    \\  \hline
        grid-I              & 4096                  & 0.25    & 0.2399 & 4.05\%        & 156.38  & 0.2425 & 0.74\%      & 322.98  \\
        grid-II             & 12856                 & 0.25    & 0.2377 & 4.93\%        & 1179.42 & 0.2428 & 0.71\%      & 2228.56 \\
        grid-III             & 4946                  & 0.25    & 0.2395 & 4.20\%        & 573.14  & 0.2428   & 0.72\%        & 725.16 
        \end{tabular}
      }
    \end{ruledtabular}
\end{table*}

\textcolor{black}{In this section, the objective is to investigate the quantitative advantages of the high-order scheme in reducing numerical dissipation. To this end, the Taylor-Green vortex with a Reynolds number of $Re = 1\,600$ is employed as a benchmark for comparison. The numerical setup is identical to that described in Section~\ref{sec:high-order codes}. The computational grids employed for the benchmark tests are presented in Fig.~\ref{fig:cell_types}, which includes the Cartesian quadrilateral grid (grid-I), the Delaunay triangular grid (grid-II), and the polygonal grid (grid-III).  The sensitivity analysis of different mesh types is presented for both solvers in Table~\ref{tab:Solver_comparison_on_grids}. The kinetic energy ($K_\mathrm{num}$) and numerical dissipation ($\varepsilon$) are compared with the analytical solutions as references. The results demonstrate that the FVMS3 scheme exhibits superior numerical accuracy compared to the traditional second-order scheme across all tested grids, with the Delaunay triangular grid (grid-II) exhibiting the most notable improvement. In addition, the computational cost of the simulation for a physical time of 10 seconds is monitored on a PC with a single CPU of 11th Generation Intel(R) Core(TM) i7-11700K, 3.60 GHz. The execution time of the FVMS3 solver is approximately $1.2\sim2.0$ times to that of the pisoFoam.
}

\section{Flow past a sphere at Re=10\,000}
\label{sec:LES_sphere}

\subsection{Description of the simulation setup}
\label{sec:model_setup}

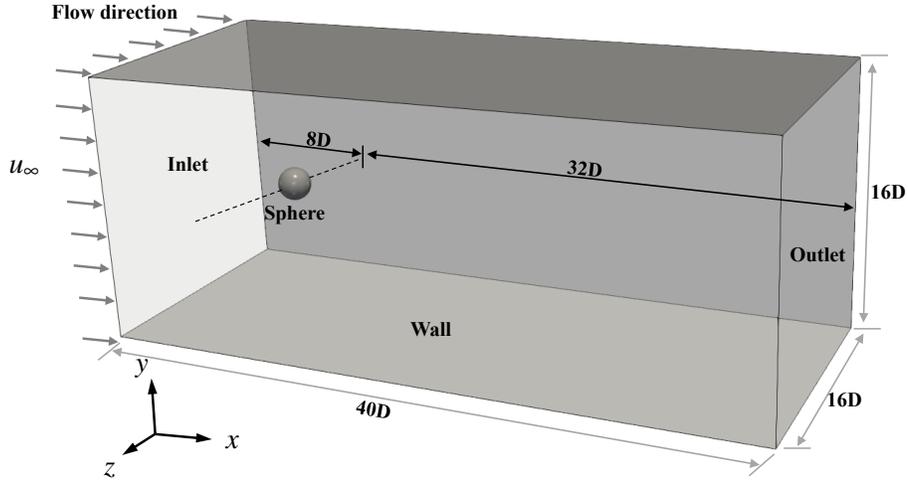
\begin{figure}[!htpb]
	\centering
	\begin{overpic}[width=12cm]{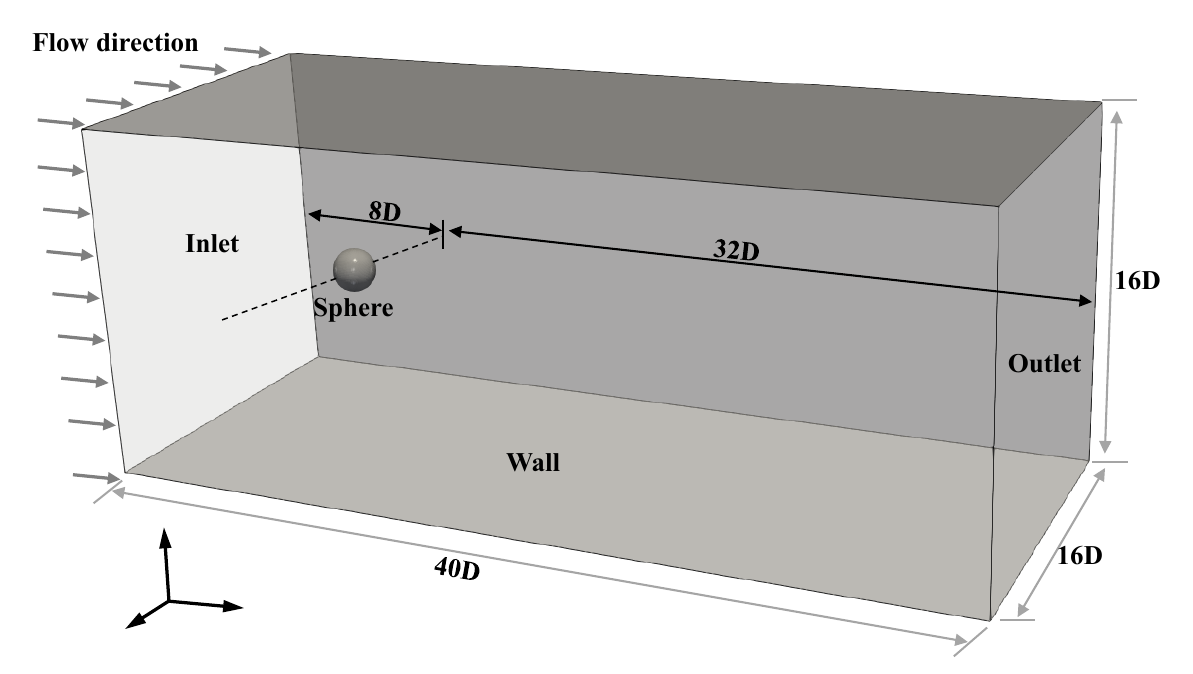}
		\put(-2,35){\color{black}{$u_{\infty}$}}
		\put(12,13){\color{black}{\textit{y}}}
		\put(8.5,1.5){\color{black}{\textit{z}}}
		\put(22,4.5){\color{black}{\textit{x}}}
	\end{overpic}		
	\caption[]{Computational domain used to simulate flow around a sphere.}
	\label{fig:computational_domain}
\end{figure}

The present study uses LES with the proposed high-order numerical schemes to simulate three-dimensional, unsteady, viscous flows around a sphere. A general illustration of the problem within our consideration is shown in Fig.~\ref{fig:computational_domain}. The setup of the computational domain is as follows: A sphere with diameter $D$ is placed on the vertical symmetry plane of a hexahedral domain. The sphere is located at a distance of $8D$ from the inlet boundary, as well as a distance of $32D$ from the downstream exit plane, which provides a relatively large space for the development of the wake flow. The spanwise length of the domain is $16D$ with the sphere located in the center, which is considered large enough to avoid undesired numerical constraints from the sidewalls. The Reynolds number $Re = u_{\infty}D/\nu$ is based on the diameter of the sphere $D$, the free stream velocity $u_{\infty}$, and the kinematic viscosity $\nu$. In addition, the boundary conditions used in the present study are similar to the DNS of Rodríguez \textit{et al.}  \cite{RODRIGUEZ2013DNS}, the LES of Yun \textit{et al.} \cite{Yun2006} and the PANS (partially averaged Navier-Stokes) of Kamble \& Girimaji\cite{Kamble2020}. At the inlet, a uniform flow velocity with a fixed value of $u_{\infty}$ is used. In addition, zero-gradient Neumann boundary conditions are applied to the far-field boundaries, and the inlet-outlet boundary condition is prescribed at the downstream plane to control the undesired backflow. In addition, the non-slip boundary condition is applied to the sphere.

\textcolor{black}{The central concern of this study is to investigate the advantages of the high-order solver with LES for solving complex turbulence problems over traditional second-order schemes which are conducted from two perspectives: (i) the ability of different numerical schemes for turbulence problems under the same set of meshes; and (ii) the accuracy and stability of higher-order schemes for solving turbulence statistics for different mesh types.} Thus, to achieve this study, the numerical setup, particularly the mesh arrangement, is similar across all numerical cases of flow past a sphere with different meshes, including hexahedral, polyhedral, and tetrahedral meshes. In all three cases, i.e. Case-hexa, Case-poly, and Case-tetra as given in Table~\ref{tab:Mesh Information}, the minimum grid size and stretching ratio of the boundary mesh were identical. The boundary mesh employs prism layers with a growth ratio of 1.025 for all three cases. The thickness of the first off-wall cell is set at $0.002D$, and approximately 20 cells are used within the boundary layer. The total number of cells in each case varies depending on the cell shape. Case-hexa has approximately 5.27 million cells, Case-poly has approximately 1.77 million cells, and Case-tetra has approximately 7.29 million cells.  The grid design and overall DOFs used in each mesh type were very different but compatible with each case.

\begin{table*}[!htpb]\small
  \caption[Mesh Information]{Details of cell types and arrangements.}
  \label{tab:Mesh Information}
    \centering
    \begin{ruledtabular}
\begin{tabular}{lccc}
Cases                                 & Case-hexa                         & Case-poly                         & Case-tetra                       \\ \hline
Cell type                             & Hexa \&   Hexa B.L. & Ploy \&   Hexa B.L. & Tetra \&   Prisms B.L. \\
Grid number                        & 5\,270\,032                      & 1\,768\,594                      & 7\,291\,802                     \\
Max. skewness                          & 0.68                         & 1.91                         & 0.72                        \\
Mesh   non-orthogonality (max,mean)   & (52.83$^\circ$,11.62$^\circ$)              & (55.34$^\circ$,12.93$^\circ$)              & (60.24$^\circ$,20.52$^\circ$)             \\
Boundary layer   mesh-first layer     & 0.002D                       & 0.002D                       & 0.002D                      \\
Boundary layer   mesh-layer count     & 20                           & 20                           & 20                          \\
Stretching   ratio                    & 1.025                         & 1.025                         & 1.025                        \\
Average number   of surfaces per cell & 6.00                         & 12.94                        & 4.10               
\end{tabular}
\end{ruledtabular}
\end{table*}

\begin{figure}[!htpb]
\centering
	\begin{overpic}[width=8cm]{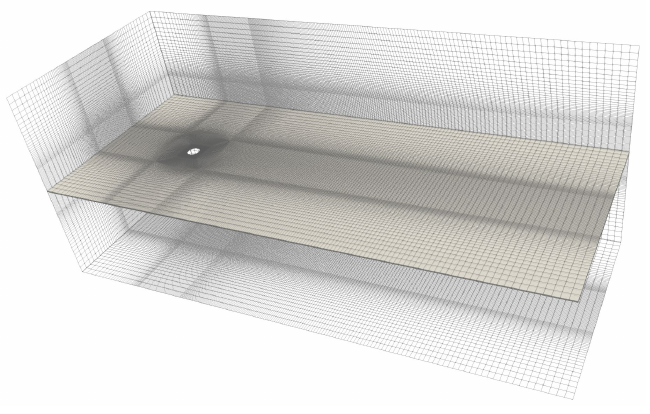}
		\put(3,60){\color{black}{(a)}}
	\end{overpic}
    \begin{overpic}[width=8cm]{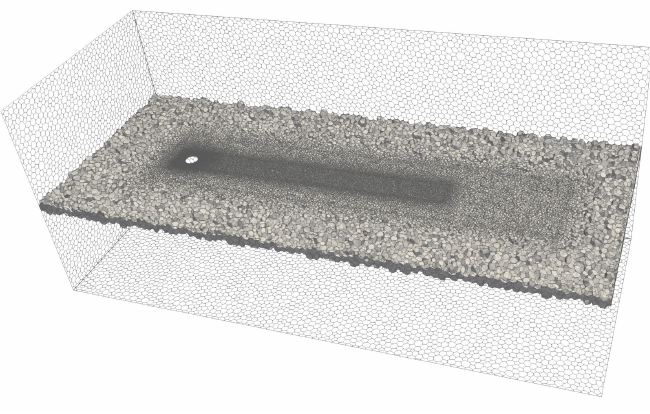}
		\put(3,60){\color{black}{(b)}}
	\end{overpic}
    \begin{overpic}[width=8cm]{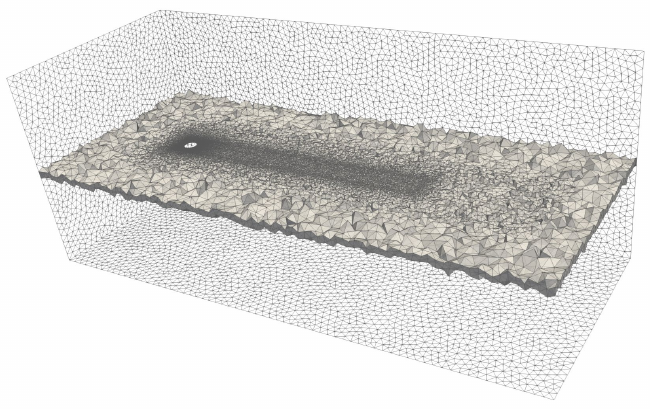}
		\put(3,60){\color{black}{(c)}}
	\end{overpic}
	\caption[]{Diagram of the overall mesh profile of different cases: (a) Case-hexa, (b) Case-poly, and (c) Case-tetra  }
	\label{fig:MeshProfile}
\end{figure}

Figure~\ref{fig:MeshProfile} presents the profiles of the mesh types and arrangements of cases for simulation of flow past a sphere. Figure~\ref{fig:MeshProfile} illustrates the computational domain, which is discretized using Cartesian background grids. In order to accurately simulate the near-wall flow dynamics of the sphere, elements near the sphere are refined within a $2D\times2D\times2D$ cube. In addition, the core region of the wake is refined with special care, while in other regions the mesh is gradually stretched towards the boundaries. A boundary layer mesh is applied around the sphere to accurately resolve boundary layer flows. In terms of near-wall resolution, the dimensionless wall normal length $y^{+}$ ($y^{+}=y u^{*} / \nu$, and $u^{*}$ the friction velocity) is limited to less than $0.7$ at $Re$  = 10\,000 by controlling the first-layer heights. Furthermore, the mesh resolution is carefully investigated by performing grid convergence studies in appendix~\ref{sec:Convergence_test}. To ensure iterative convergence while solving the governing equations, the maximum Courant-Friedrichs-Lewy (CFL) number is limited to $0.3$ throughout all simulations. To ensure the reliability of the obtained statistics, $900D/u_{\infty}$ time units (about 20 flow-throughs, the time required for the mean flow to travel through the domain) are first simulated until a ``statistically stationary state'' of the turbulent flow develops, which is reported to be sufficient for the subcritical Reynolds number with respect to flow past a sphere \citep{Constantinescu2004LES}. It should be noted that after the initial simulation of $900D/u_{\infty}$ time units, $t^{*} = 0$ ($t^{*} = tu_{\infty}/{D}$) is defined and another $200D/u_{\infty}$ time units (about 40 vortex shedding cycles of the dominant frequency) are used to collect data for analysis. In order to facilitate the convergent solution of the flow, a criterion of $1\times10^{-8}$ was set for the maximum difference between two iteration steps of the physical quantities (i.e., $\mathbf{u}$ and $p$). In the present study, more than $210\,000$ CPU hours were used to run simulations and collect statistics on a 1280-core computer cluster.

\subsection{Instantaneous flow structures}\label{sec:flow_structures}
\begin{figure*}[!htpb]
  \centering
  \begin{minipage}[c]{1\linewidth}
  \centering
  \begin{overpic}[width=13cm]{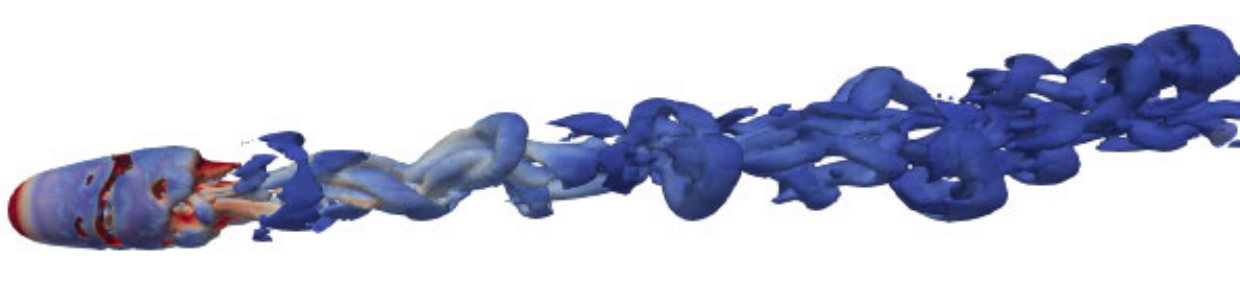}
  \put(-3,14){\color{black}{(a)}}
  \end{overpic}
    \begin{overpic}[width=13cm]{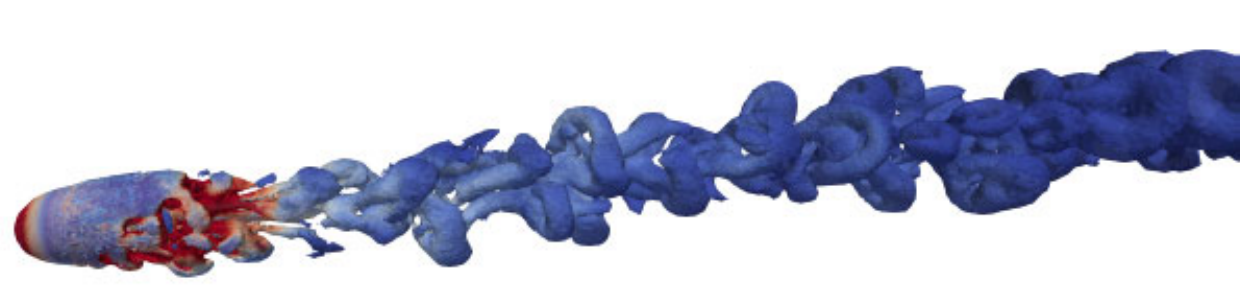}
  \put(-3,14){\color{black}{(b)}}
  \end{overpic}
    \begin{overpic}[width=13cm]{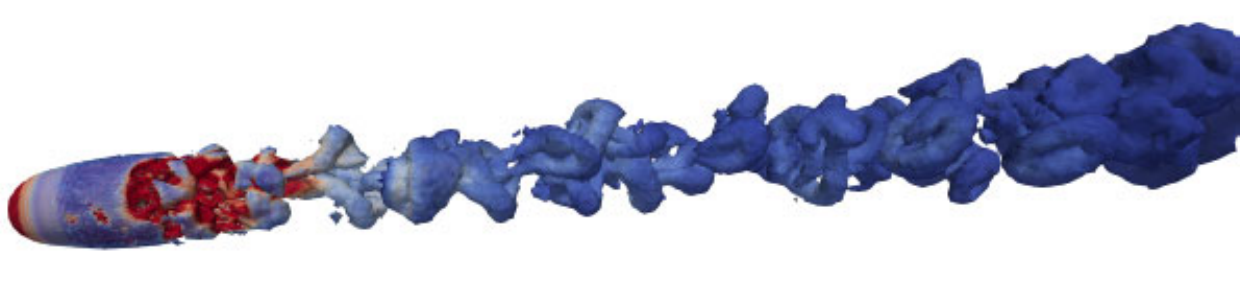}
  \put(-3,14){\color{black}{(c)}}
  \put(85,2){\includegraphics[scale=0.2]{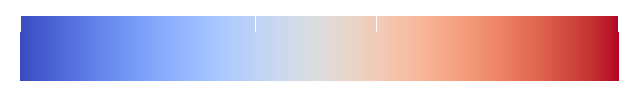}}
  \put(85,5){\color{black}{0}}
  \put(100,5){\color{black}{190}}
  \put(90,5){\color{black}{$\omega D/ u_\infty$}}
  \end{overpic}
  \end{minipage}
	\caption[]{Vortex structures of turbulent flow around a sphere at $Re$  = 10\,000 solved by FVMS3, using the iso-surface of the \textit{Q} criterion with $QD^2/u_{\infty}^2 = 0.001$ at $t^* = 160$, and coherent structures are colored by the nondimensional instantaneous vorticity magnitude from 0 to 190, (a) Case-hexa; (b) Case-poly; (c) Case-tetra.}
	\label{fig:vortex_structures_fvms3}
\end{figure*}

\begin{figure*}[!htpb]
  \centering
  \begin{minipage}[c]{1\linewidth}
  \centering
  \begin{overpic}[width=13cm]{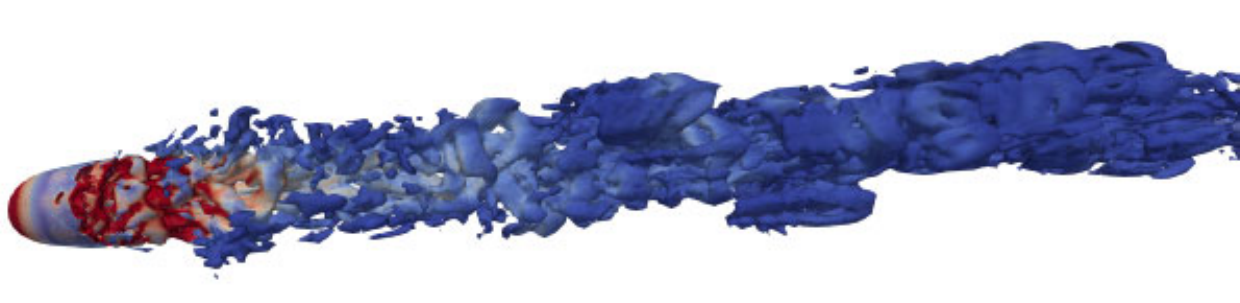}
  \put(-3,14){\color{black}{(a)}}
  \end{overpic}
    \begin{overpic}[width=13cm]{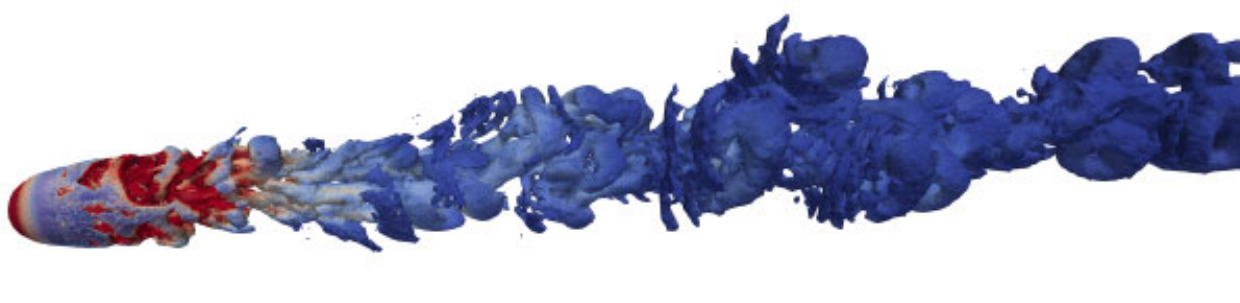}
  \put(-3,14){\color{black}{(b)}}
  \end{overpic}
    \begin{overpic}[width=13cm]{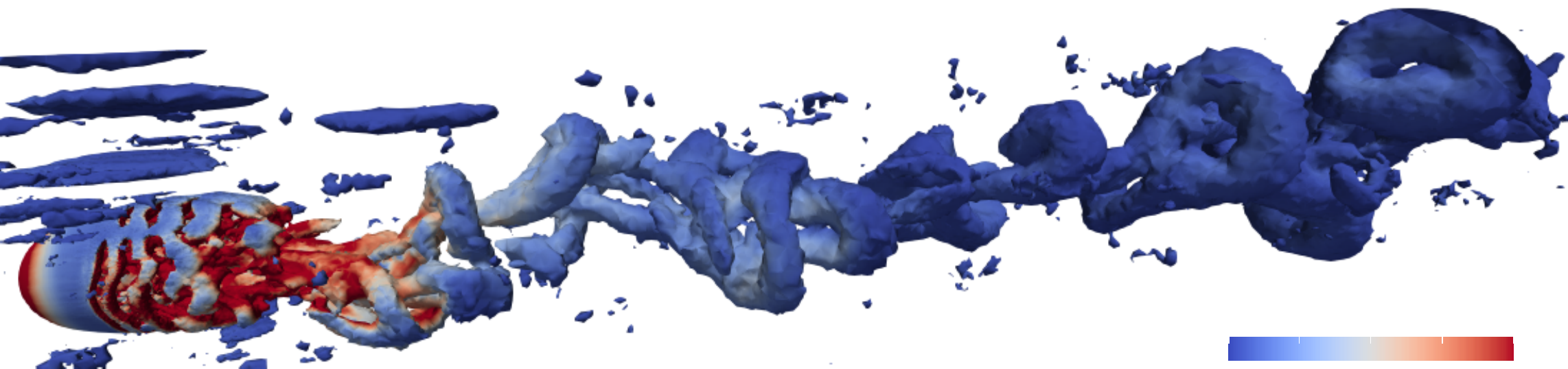}
  \put(-3,14){\color{black}{(c)}}
  \put(78,3){\color{black}{0}}
  \put(94,3){\color{black}{190}}
  \put(83,3){\color{black}{$\omega D/ u_\infty$}}
  \end{overpic}
  \end{minipage}
    \caption[]{Same as Fig.~\ref{fig:vortex_structures_fvms3} but solved by pisoFoam, (a) Case-hexa; (b) Case-poly; (c) Case-tetra.}
	\label{fig:vortex_structures_pisoFoam}
\end{figure*}
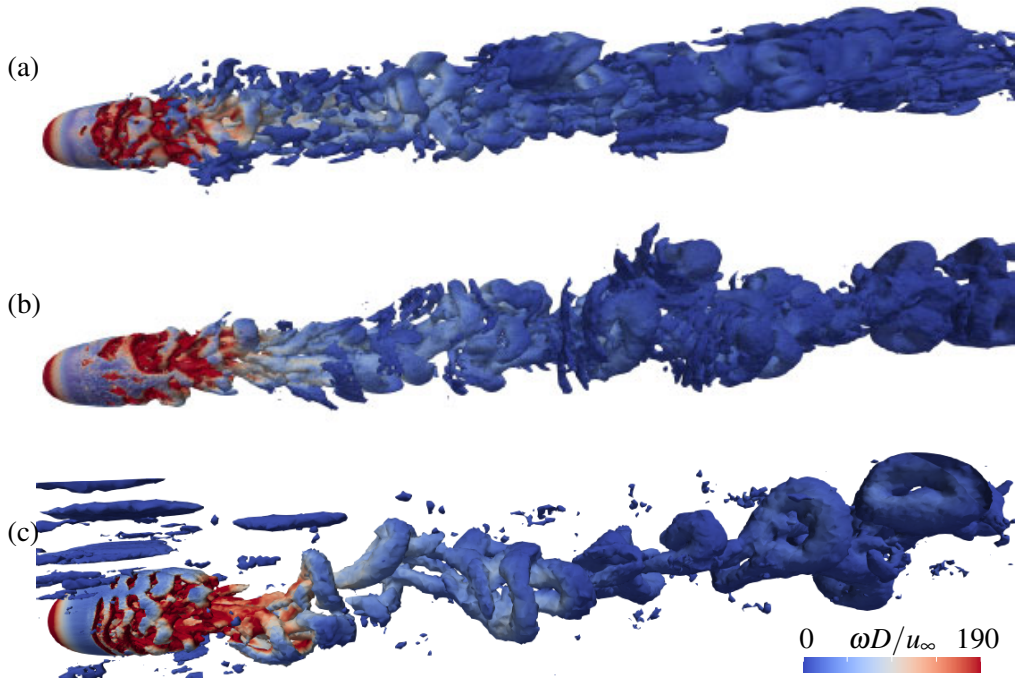

Figures~\ref{fig:vortex_structures_fvms3} and \ref{fig:vortex_structures_pisoFoam} show the instantaneous structures of vortices around a sphere at $Re$  = 10\,000 identified by the \textit{Q}-criterion\cite{hunt1988eddies}. As the flow passes through the sphere at a subcritical Reynolds number, the flow field is characterized by the transition of the boundary layer flow from laminar to turbulent and the unsteady shedding of vortex structures. In addition, hairpin-like structures are observed just beyond the recirculation region as the vortices alternately separate into the wake. As shown in Fig.~\ref{fig:vortex_structures_fvms3}, the vortex leg and head of the shedding hairpin as well as the K-H instability are accurately predicted by the high-order FVMS3 scheme with LES. The instantaneous flow structures predicted by the high-order FVMS3 scheme for different grid types are consistent with the results of Constantinescu \& Squires\cite{Constantinescu2004LES}. These results suggest that the present model can produce high-fidelity numerical solutions regardless of the type of grid element. However, as shown in Fig.~\ref{fig:vortex_structures_pisoFoam}, the conventional second-order scheme predicts vortical structures that are relatively disordered due to numerical noise, which degrades the solution quality. The results of the low-order numerical model simulation demonstrate that numerical noise overwhelms the physical vortex structure due to severe numerical dissipation, and the flow fields simulated by pisoFoam are more sensitive to mesh types. For the tetrahedral mesh (Case-tetra) presented in Fig.~\ref{fig:vortex_structures_pisoFoam}(c), there are numerous non-physical structures in the flow field due to numerical noise rather than flow physics.

\subsection{Mean flow topology}\label{sec:mean_flow_topology}
\subsubsection{Flow statistics around the sphere}

\begin{table}[!htpb]\small
    \caption{Comparison of statistical parameters simulated by FVMS3 and pisoFoam for flow past a sphere at $Re$  = 10\,000 with available references. The statistics are the Strouhal number of vortex shedding $\mathrm{St} = f_{vs} D/{u_\infty}$, the separation angle $\phi_s(^\circ)$, and the drag coefficient $C_D = F_D/(\frac{1}{2} \rho {u_{\infty}}^2 A)$. Note that the Reynolds numbers of all cases used as references are $10\,000$.}
	\label{tab:Mean_properties}
    \centering
    \begin{ruledtabular}
\begin{tabular}{lcccc}

                           & Case                             & $\mathrm{St}$    & $\phi_s(^\circ)$ & $C_D$                 \\ \hline
\multirow{3}{*}{\textcolor{black}{Present: FVMS3}}   & Case-hexa         & 0.192 & 86.56       & 0.401                            \\ 
                           & Case-poly                                 & 0.190 & 85.94       & 0.403                             \\ 
                           & Case-tetra                                & 0.190 & 85.55       & 0.395                              \\ [3pt] \hline 
                           
\multirow{3}{*}{Low-order model: pisoFoam}   & Case-hexa                      & 0.158 & 88.51       & 0.407                            \\
                           & Case-poly                                 & 0.175 & 87.13       & 0.415                             \\
                           & Case-tetra                                & 0.174 & 88.17       & 0.424                                \\[3pt] \hline 
\multirow{4}{*}{Reference} 
			& LES\citep{Yun2006}     & 0.17 & 90       & 0.393         \\ 
			& DNS\citep{RODRIGUEZ2013DNS}         & 0.195 & 84.70       & 0.402             \\
			& LES\citep{Constantinescu2004LES}     & 0.195 & 84-86       & 0.393        \\ 
			& Exp.\citep{achenbach_1972,achenbach_1974}  & 0.195 & 82.50       & 0.400 \\
\end{tabular}
    \end{ruledtabular}
\end{table}

In this section, we compare typical flow statistics in the near wake of a sphere with existing references \citep{RODRIGUEZ2013DNS,Constantinescu2004LES}. As shown in Table~\ref{tab:Mean_properties}, the drag coefficients obtained from simulations of three different mesh types employing the current high-order scheme exhibit favorable agreement with previous results. The maximum relative difference, for the results from Case-tetra, does not exceed $0.25\%$ when referencing the DNS results reported by Rodríguez \textit{et al.}  \cite{RODRIGUEZ2013DNS}. Moreover, the characteristics of the recirculation bubble in our simulation compare well with those in the previous studies. Regarding the separation angle, a laminar boundary layer separation of about $86^\circ$ is also accurately captured by the present LES. On the other hand, for the present different mesh types, however, the $\mathrm{St}$ results obtained by the pisoFoam solver are smaller than the reference solutions. More quantitatively, the relative differences in the Strouhal number of vortex shedding are calculated similarly, taking the DNS results as a reference. The $\mathrm{St}$ is underestimated by pisoFoam for the results from Case-hexa with the error of $-18.97\%$ which is higher than that of FVMS3 ($-1.54\%$) from Case-hexa. The results indicate that the present solver is one of the attractive choices for the accurate and robust simulation of turbulent flows around the sphere. This shows that the results of the present scheme are less dependent on the type of grid elements, which provides significant advantages over traditional low-order schemes for solving complex high-Reynolds-number turbulence problems.

\begin{figure*}[!htpb]
  \centering
  \begin{minipage}[c]{1\linewidth}
  \centering
  \begin{overpic}[width=8cm]{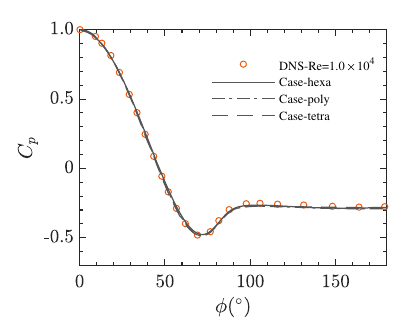}
  \put(3,70){\color{black}{(a)}}
  \end{overpic}
    \begin{overpic}[width=8cm]{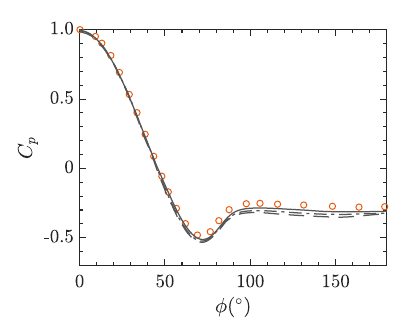}
  \put(3,70){\color{black}{(b)}}
  \end{overpic}
  \end{minipage}
    \begin{minipage}[c]{1\linewidth}
  \centering
  \begin{overpic}[width=8cm]{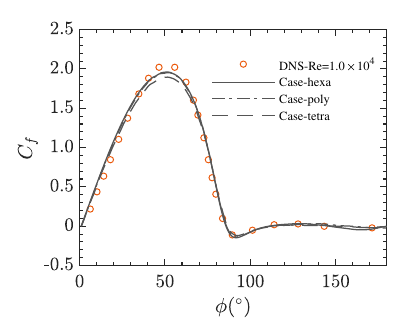}
  \put(3,70){\color{black}{(c)}}
  \end{overpic}
    \begin{overpic}[width=8cm]{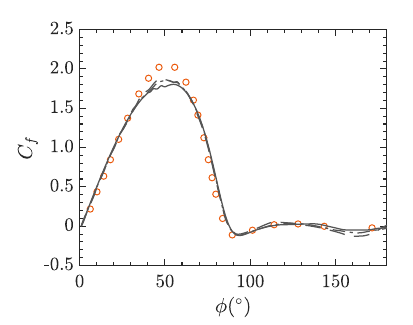}
  \put(3,70){\color{black}{(d)}}
  \end{overpic}
  \end{minipage}
    \caption[]{Angular distribution of (a,b) mean pressure coefficient and (c,d) mean skin-friction coefficient around the sphere at $Re$  = 10\,000. Results from present solver: FVMS3 (a,c) and open-source solver: pisoFoam (b,d); compared to DNS results by Rodríguez \textit{et al.}  \cite{RODRIGUEZ2013DNS} } 
	\label{fig:Cp_Tau}
\end{figure*}

As shown in Fig.~\ref{fig:Cp_Tau}(a), with respect to the angular distribution of the mean pressure coefficient ($C_p = ({p-p_\infty})/{\frac{1}{2} \rho {u_{\infty}}^2}$), an excellent agreement is found for the present results computed by FVMS3 scheme compared to the corresponding DNS solutions in Rodríguez \textit{et al.}  \cite{RODRIGUEZ2013DNS} while some of the highest discrepancies are observed for pisoFoam in the region of $\phi > 100^\circ$, where the boundary layer separation occurs. In addition, we also report another classical verification parameter, the skin friction coefficient ($C_f = \tau_w/\rho u_{\infty}^2 Re^{0.5}$), to evaluate the capability of numerical schemes in predicting the wall shear stress. As shown in Fig.~\ref{fig:Cp_Tau}(c), our results are in good agreement with the LES solutions by Constantinescu \& Squires\cite{Constantinescu2004LES}, and DNS solutions by Rodríguez \textit{et al.}  \cite{RODRIGUEZ2013DNS}, indicating that the variation of the skin friction coefficient along the sphere is well reproduced by the present scheme. For pisoFoam, the magnitude of skin friction seems to be slightly lower than our results as well as the reference solutions, especially in the turbulent boundary layer region. In addition, the separation of the laminar boundary layer occurs near the equator of the sphere at $85^\circ$, which is well identified by the distribution of the nondimensional skin friction coefficient, similar to those observed in the experiments by Achenbach\cite{achenbach_1972}.

\subsubsection{Flow statistics in the near wake}

\begin{figure*}[!htpb]
  \centering
  \begin{minipage}[c]{1\linewidth}
  \centering
  \begin{overpic}[width=8cm]{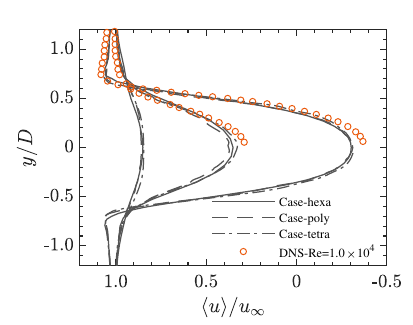}
  \put(3,70){\color{black}{(a)}}
  \put(75,53){\scriptsize{\color{black}{$x/D=1.6$}}}
  \put(60,45){\scriptsize{\color{black}{$x/D=2.5$}}}
  \put(36,40){\scriptsize{\color{black}{$x/D=5.0$}}}
  \end{overpic}
    \begin{overpic}[width=8cm]{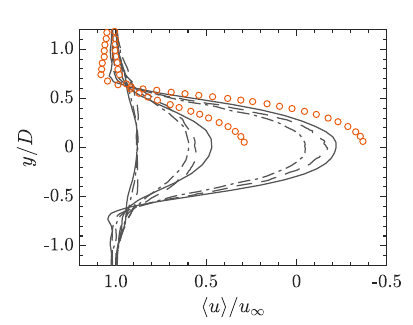}
  \put(3,70){\color{black}{(b)}}
  \end{overpic}
  \end{minipage}
  \caption[]{Near wake flow statistics: time-averaged streamwise velocity at $x/D = 1.6,\ 2.5\ \mathrm{and}\ 5.0$ at $Re$  = 10\,000 compared to DNS results of Rodríguez \textit{et al.}  \cite{RODRIGUEZ2013DNS} (a) Results from present solver: FVMS3; (b) Results from open-source solver: pisoFoam; }
 \label{fig:mean_velocity}
\end{figure*}

\begin{figure*}[!htpb]
\centering
\subfigure{  
  \begin{overpic}[height=6.3cm]{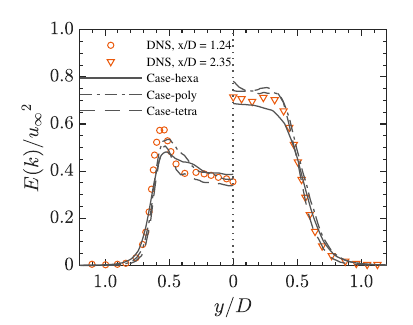}
  \put(3,70){
  \color{black}{(a)}}
  \end{overpic}
  \begin{overpic}[height=6.3cm]{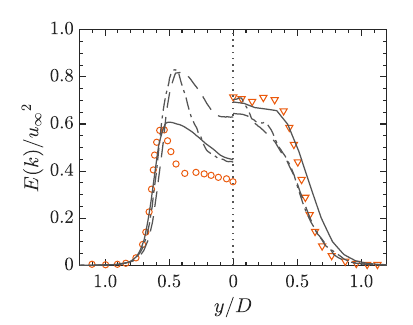}
  \put(3,70){\color{black}{(b)}}
  \put(65,63){\footnotesize{\color{black}{$x/D=2.35$}}}
  \put(25,63){\footnotesize{\color{black}{$x/D=1.24$}}}
  \end{overpic}}
\caption{Comparison of mean turbulent kinetic energy in the near wake solved by the present high-order solver for different grid elements at measurement locations $x/D = 1.24$ (left column) and $x/D = 2.35$ (right column); Solved by (a) the present high-fidelity model FVMS3 and  (b) the open-source solver: pisoFoam. Reference data identified as circle and triangular: DNS results from Rodríguez \textit{et al.}  \cite{RODRIGUEZ2013DNS} .}
\label{fig:wakeTKE}
\end{figure*}

The time-averaged streamwise velocity profiles ($\langle u \rangle /u_{\infty}$) at different downstream locations in the wake ($x/D = 1.6,\ 2.5,\ \mathrm{and}\ 5.0$) are shown in Fig.~\ref{fig:mean_velocity}. The ``U'' shaped profile of the mean velocity inside the recirculation bubble is well illustrated by the present simulation which shows good agreement with reference DNS solutions. Note that all the present simulations show a smaller velocity deficit, while the FVMS3 scheme shows a closer agreement with DNS results than the pisoFoam solver. Furthermore, when considering cases with different mesh types, the FVMS3 method shows better agreement with DNS solutions compared to pisoFoam. It also suggests that the present model predicts a slower recovery in the recirculation region, and similar observations have also been reported by Kamble \& Girimaji \cite{Kamble2020} for the study of the wake of a sphere at subcritical Reynolds number ($Re = 3\,700$) using PANS simulation. The comparison of mean turbulent kinetic energy (TKE) in the near wake solved by the present high-order solver and the traditional low-order solver for different grid elements is shown in Fig.~\ref{fig:wakeTKE}. The solver based on FVMS3 accurately predicted the general profile in the near wake despite different types of meshes. However, the solver based on a traditional low-order scheme was unable to accurately predict the TKE, particularly in the recirculation region.

\subsection{spectrum characteristics}
\label{sec:spectrum_characteristics}

\begin{figure*}[!htpb]
	\centering
    \begin{minipage}[c]{1\linewidth}
		\centering
		\begin{overpic}[width=10cm]{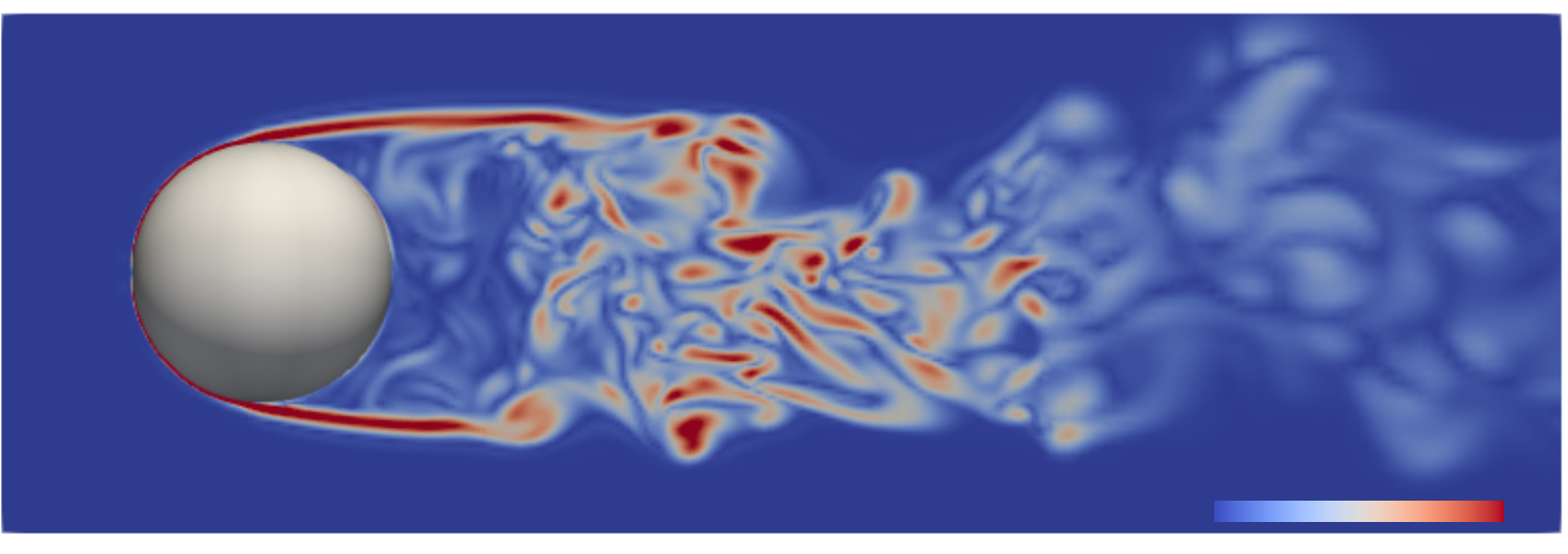}
			\put(-6,35){\color{black}{(a)}}
            \put(33,27){\color{green}{$\circ$}}
            \put(33,30){\color{green}{$P_1$}}
            \put(81,27){\color{green}{$\circ$}}
            \put(81,30){\color{green}{$P_2$}}
            \put(83,5){\scriptsize\color{white}{$\omega D/u_{\infty}$}}
            \put(76,5){\scriptsize\color{white}{0}}
            \put(94,5){\scriptsize\color{white}{10}}
		\end{overpic}
	\end{minipage}
    \begin{minipage}[c]{1\linewidth}
		\centering
		\begin{overpic}[width=7.5cm]{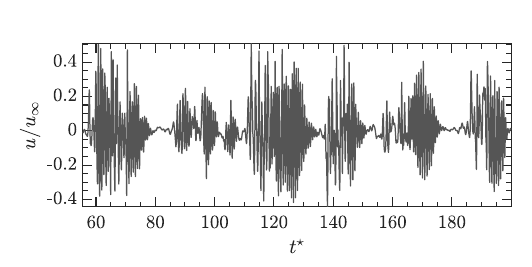}
			\put(3,40){\color{black}{(b)}}
		\end{overpic}
		\begin{overpic}[width=7.5cm]{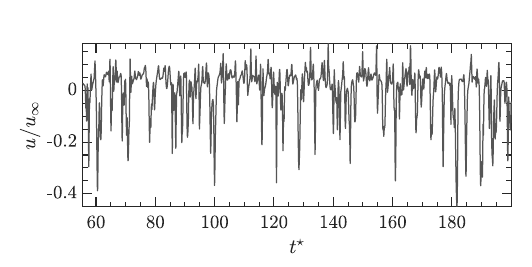}
			\put(3,40){\color{black}{(c)}}
		\end{overpic}
	\end{minipage}
	\begin{minipage}[c]{1\linewidth}
		\centering
		\begin{overpic}[width=7.5cm]{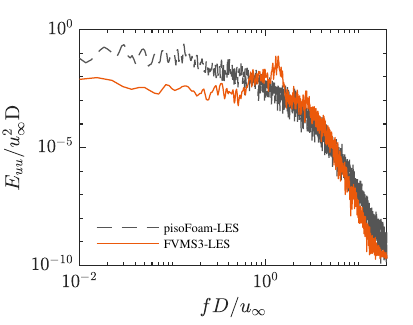}
			\put(3,70){\color{black}{(d)}}
			\put(68,67){\color{blue}{$f_{KH}$}}
            \put(66,63.4){\color{blue}{$\circ$}}
		\end{overpic}
		\begin{overpic}[width=7.5cm]{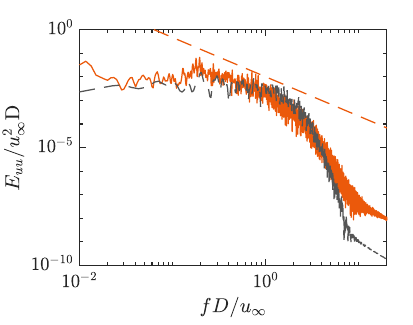}
			\put(3,70){\color{black}{(e)}}
			\put(68,63){\color{red}{-5/3}}
            \put(50,67){\color{blue}{$f_{vs}$}}
            \put(46.5,62.8){\color{blue}{$\circ$}}
		\end{overpic}
	\end{minipage}
	\caption[]{(a) Schematic of the instantaneous vorticity field; (b,c) Time history of the streamwise velocity simulated by FVMS3 with LES for Case-Hexa; (d,e) One-dimensional power spectrum of the streamwise velocity at locations $P_1:\ x/D = 1. 0,\ r/D = 0.6$ (b,d) and $P_1:\ x/D = 4.0,\ r/D = 0.6$ (c,e), simulated by the FVMS3 and pisoFoam with LES, where $r$ is the radial distance between the sampling points and the wake center line.}
	\label{fig:powerSpectrum_FVMS3}
\end{figure*}
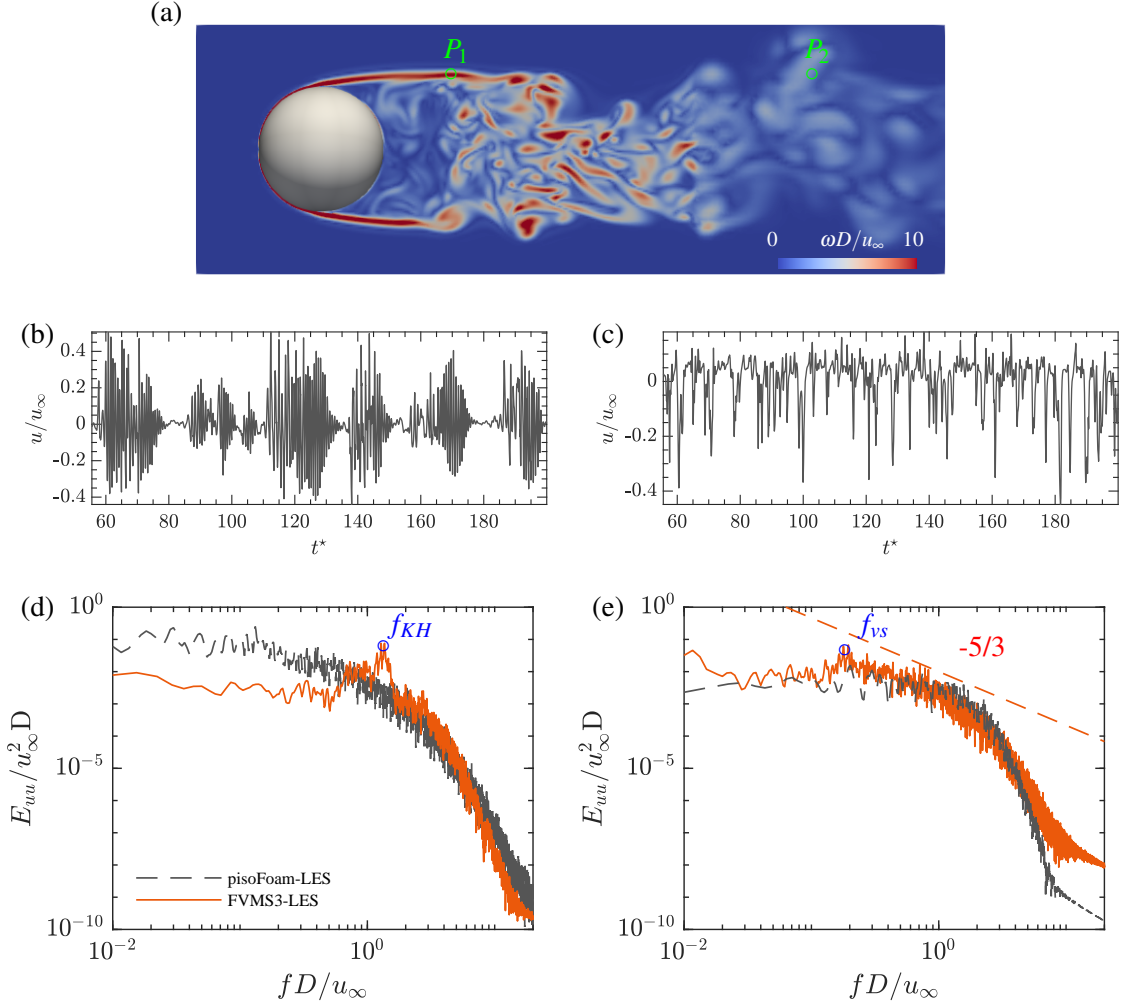

\textcolor{black}{The success of numerical schemes used for turbulent simulation depends upon the ability to identify key flow mechanisms responsible for the generation of coherent structures. For flow past a sphere at a subcritical Reynolds number, two frequency modes of instabilities were reported \cite{Kim1988, Constantinescu2004LES, rodriguez2011, RODRIGUEZ2013DNS}: (i) a low-frequency mode associated with the large-scale, i.e., vortex-shedding, instability of the wake, and (ii) a high-frequency mode associated with the small-scale instabilities of the separating shear layer.} Thus, for an in-depth exploration of the capability of the numerical solver in predicting these key flow physics, we have further investigated the spectral behavior of the flow past a sphere at $Re = 10\,000$ by employing Welch's power spectral density distributions \citep{Welch1967} of the streamwise velocity at various measurement locations. The flow characteristics of a sphere wake within this range of Reynolds number ($Re$) exhibit three distinct shear layers characterized by different flow physics. These include the laminar boundary layer, the free shear layer, and the wake flow. For flow past a sphere, the laminar boundary layer separates from the sphere and generates the free shear layers and a strong recirculation region in the near wake. In the free shear layer, the Kelvin-Helmholtz rollers are generated due to the Kelvin-Helmholtz instability. In the wake flow, hairpin-like vortex structures are generated just beyond the recirculation region and the vortices alternately shed into the far wake. The relevance of this coherent structure has prompted a multitude of studies over the past several years. \citep{RODRIGUEZ2013DNS,rodriguez2011,Pereira2018SRS}. 

In this section, we focused on the spectral results of Case-Hexa from both solvers for comparison, for which the pisoFoam performs better on the hexahedra mesh according to the above results. In Figure~\ref{fig:powerSpectrum_FVMS3}(a), a flow visualization of the instantaneous vorticity field and a schematic of two probes are presented. In addition, it is now feasible to visually observe the Kelvin-Helmholtz rollers, evident in both the time history of velocity fluctuations and the corresponding spectrum, similar to that observed by Pereira \textit{et al.} (2018)\cite{Pereira2018SRS}. The first probe, situated at $x/D = 1.0,\ r/D = 0.6$, is strategically placed to investigate the Kelvin-Helmholtz instability in the free shear layer. Meanwhile, the second probe is positioned in the near-wake of the sphere at $x/D = 4.0,\ r/D = 0.6$ to estimate the shedding frequency. As depicted in Figs.\ref{fig:powerSpectrum_FVMS3}(b) and \ref{fig:powerSpectrum_FVMS3}(d), the obtained energy spectrum demonstrates the presence of the K-H instability, which manifests as intermittent bursts of high-frequency activity within the velocity field, characterized by the frequency $f_{KH}$. Moreover, for the second probe, the vortex shedding frequency $f_{vs} = 0.192$ in the energy spectra is captured by the present LES, as illustrated in Fig.\ref{fig:powerSpectrum_FVMS3}(e). However, it is noteworthy that this instability, although less pronounced, persists and remains visible in the spectrum, as also reported in the DNS study of flow past a sphere at the same Reynolds number \citep{RODRIGUEZ2013DNS}. The diminished vortex-shedding signature may be attributed to (i) the necessity of a large sampling period for accurately measuring such low-frequency values and (ii) the observed phenomenon that, in the case of the sphere at the present subcritical Reynolds number, the vortex-shedding seems to exhibit lower amplitude, indicating lower energy compared to other bluff bodies\citep{RODRIGUEZ2013DNS}. Furthermore, the present results indicate that from the velocity spectrum, $f_{vs}$, exhibits a pronounced peak, while $f_{KH}$ displays a relatively broadband, which is consistent with the results of flow past a cylinder at a subcritical Reynolds number \citep{Prasad1997Instability, Pereira2018SRS}. Moreover, the averaged spectrum in Fig.~\ref{fig:powerSpectrum_FVMS3}(e) reveals a clear inertial subrange over a few octaves, consistent with the $-5/3$ law, in accordance with the well-known ``K41 theory''. \textcolor{black}{Based on the spectrum analysis in this subsection and the vortex structures presented in Figs.~(\ref{fig:vortex_structures_fvms3}) and (\ref{fig:vortex_structures_pisoFoam}), results show that the present high-order numerical model adeptly identifies these key flow physics for flow past a sphere such as two frequency modes of instabilities which are important for the generation of flow structures. However, the low-order scheme was unable to accurately predict this instability, which is less pronounced.}

\section{Conclusion}
\label{sec:conclusions}
The objective of this study is to identify the utility and effectiveness of the high-order scheme for simulating unsteady turbulent flows in comparison to traditional second-order schemes. To achieve this objective, the investigations are conducted from two perspectives: (i) the ability of different numerical schemes for turbulence problems under the same set of meshes; and (ii) the accuracy and stability of higher-order schemes for solving turbulence statistics for different mesh types of hexahedral, tetrahedral, and polyhedral cells. To address this issue, we develop a numerical model for incompressible flows by solving Navier-Stokes equations with the third-order FVMS3 scheme \cite{Xie2019High-fidelitysolver,Xie2020consistent} for the spatial discretization. The numerical simulation is also carried out with pisoFoam for comparisons.

The decay of the Taylor-Green vortex at $Re=100,\,1\,600$ are considered for verification, in which analytical solutions are available. In this case, we evaluate the performance of the present model in terms of numerical accuracy, convergence behavior, energy conservation, and numerical dissipation. For the TGV case, compared with pisoFoam with the conventional scheme, the present model significantly improves the numerical accuracy with convergence rates of more than 2.5. More importantly, the numerical dissipation of the present model is nearly 1/10 of pisoFoam, which is a distinct advantage in LES since it might overwhelm the effect of the turbulent viscosity model for high-Reynolds number flows. Moreover, the comparison is extended to include solvers of different orders, thereby illustrating the superiority of the proposed method across a wider range of alternatives. The two other solvers, i.e. DINO and Nek5000, exhibited superior accuracy for this problem, as they employ higher-order schemes for spatial discretization. However, these solvers require regular hexahedral meshes, which restricts their potential applications. It is encouraging to note that the present scheme is capable of achieving high-order numerical accuracy regardless of the type of grid used, which is a significant advantage for practical applications involving complex geometries. Subsequently, we proceed to examine the quantitative merits of the high-order scheme in curbing numerical dissipation, with a particular focus on mesh types. The results demonstrate that the FVMS3 scheme exhibits superior numerical accuracy in comparison to the traditional second-order scheme across all tested grids, with the Delaunay triangular grid exhibiting the most notable improvement.

After verification, we investigated the flow around a sphere at a diameter-based Reynolds number of $10\,000$ utilizing LES, which encompasses a multitude of intricate flow characteristics. A meticulous analysis was conducted of the instantaneous flow structures and mean flow topology, including the flow statistics around the sphere and in the near wake, and spectrum characteristics. The principal findings and results are as follows:

\begin{enumerate}
    \item For the vortex structures, the high-order scheme with LES accurately predicts the vortex leg and head of the shedding hairpin, as well as the K–H instability, for different grid types that are consistent with the results of Constantinescu \& Squires \cite{Constantinescu2004LES}. However, the conventional second-order scheme is more sensitive to mesh types, which results in numerous non-physical structures in the flow field due to numerical noise rather than flow physics, especially for the tetrahedral cells. 
    
     \item For the prediction of typical flow statistics of the sphere, including hydrodynamic forces, surface skin friction, vortex shedding, boundary layer separation, and wake TKE, the present solver provides significant advantages over traditional low-order schemes. Typically, the vortex shedding frequency is underestimated by pisoFoam for the results from Case-hexa with the error of $-18.97\%$ which is higher than that of FVMS3 ($-1.54\%$). Moreover, for high-order statistics, the traditional solver was unable to accurately predict the TKE, particularly in the recirculation region.
     
     \item For the spectrum characteristics predicted by two different solvers on hexahedral mesh, the obtained energy spectrum by FVMS3 captures the Kelvin-Helmholtz instability with a characteristic frequency $f_{KH}$ and the vortex shedding frequency $f_{vs} = 0.192$ in the energy spectra. \textcolor{black}{The results demonstrate that the present high-order numerical model is highly effective in identifying the critical flow physics associated with flow past a sphere. This includes the identification of two distinct frequency modes of instability, which play a pivotal role in the generation of flow structures. However, the low-order scheme exhibited limitations in accurately predicting this particular instability, which is less pronounced.}
\end{enumerate}

\textcolor{black}{
It is noteworthy that this numerical solver has been carefully validated and successfully applied to a diverse range of complex problems in ship and ocean engineering, such as wave propagation and runup \cite{Huang2021High-fidelity}, plunging wave breaking at \(\mathrm{Re} = 3.57 \times 10^6\) \cite{Jiang2022WaveBreaking}, and the turbulence and hydrodynamic noise for an axisymmetric hull at \(\mathrm{Re} = 1.2 \times 10^7\) \cite{Jiang2024SUBOFFhull}. In conclusion, this study offers substantial evidence in support of the effectiveness of the high-order scheme for numerical simulation of unsteady turbulent flows.}

\textcolor{black}{As a final closing comment to this study, here, we provide a summary of the key advantages and limitations of the high-order method for turbulence simulation for the sake of further exploration in the field. Key advantages include (i) The proposed high-order numerical model can significantly reduce the numerical dissipation which is of critical importance for the LES simulation of high-Reynolds number flows, as the numerical dissipation may potentially overwhelm the impact of the turbulent viscosity model within the context of LES. (ii) The present high-order method shows less dependence on mesh types, breaking the dependence of traditional high-order methods on the quality and type of mesh, which is appealing for the simulation of problems with complex geometry. (iii) The proposed high-order numerical model can significantly reduce the numerical noise which is appealing for the simulation of computational hydro- and aeroacoustics as well as high Reynolds number turbulent flow due to these problems being sensitive to numerical noise \cite{Wang2006noiseReview}. On the other hand, the computational cost is higher for high-order schemes. The present model takes approximately twice the computational time as the conventional one on the same grid resolution if one does not consider the improvements in numerical accuracy and solution quality. Thus, further advances in numerical efficiency of high-order methods for high Reynolds number turbulent flow give hope for progress in this area.}

\begin{acknowledgments}
The authors are grateful to Professor S.L. Tang from Harbin Institute of Technology for the fruitful discussions that inspired this work. The authors also acknowledge computational resources from the Center for High-Performance Computing Center at Shanghai Jiao Tong University. This work was supported in part by Shanghai Rising-Star Program (No. 23QA1405000), the fund from Shanghai Pilot Program for Basic Research - Shanghai Jiao Tong University (No. 21TQ1400202), as well as the fund from the Fundamental Research Funds for the Central Universities and the support from National Natural Science Foundation of China (No. 91752104).
\end{acknowledgments}
\section*{Author declarations}
The authors declare that they have no conflict of interest.
\section*{Data availability}
The data that support the findings of this study are available from the corresponding author upon reasonable request.
\section*{Author ORCIDs}
\noindent Peng Jiang, \href{https://orcid.org/0000-0002-9072-8307}{https://orcid.org/0000-0002-9072-8307};\\
	Yichen Huang, \href{https://orcid.org/0000-0002-6603-5156}{https://orcid.org/0000-0002-6603-5156};\\
	Yong Cao, \href{https://orcid.org/0000-0002-0262-8251}{https://orcid.org/0000-0002-0262-8251};\\
	Shijun Liao, \href{https://orcid.org/0000-0002-2372-9502}{https://orcid.org/0000-0002-2372-9502};\\
	Bin Xie, \href{https://orcid.org/0000-0002-4218-2442}{https://orcid.org/0000-0002-4218-2442}.
\appendix
\color{black}
\section{Grid convergence test for Case-hexa}\label{sec:Convergence_test}

\begin{figure}[!htpb]
	\centering  
	\begin{minipage}[l]{1\linewidth}
		\centering
		\begin{overpic}[width=8cm]{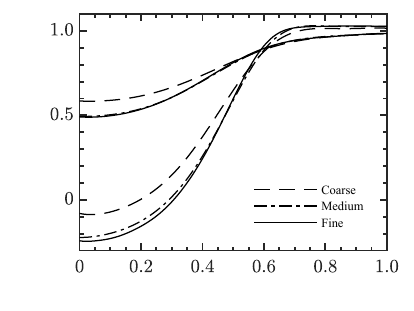}
			\put(5,70){\color{black}{(a)}}
			\put(21,35){\footnotesize\color{black}{$x/D = 1.6$}}
			\put(21,58){\footnotesize\color{black}{$x/D = 2.5$}}
			\put(53,3){\footnotesize\color{black}{$r/D$}}
			\put(3,33){\color{black}{\rotatebox{90}{$\langle u \rangle /u_{\infty}$}}}
		\end{overpic}
		\begin{overpic}[width=8cm]{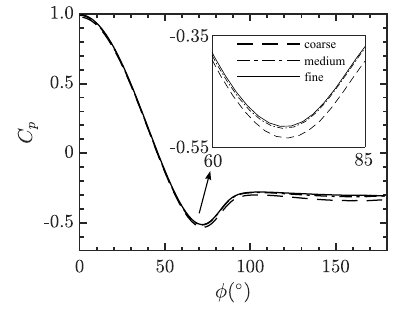}
			\put(2,70){\color{black}{(b)}}
		\end{overpic}
	\end{minipage}
	\caption[]{Comparison of typical physical quantities from three different mesh resolutions for the hexahedra mesh, i.e. coarse, medium, and fine grids (Table~\ref{tab:mesh_convergence}): (a) the time-averaged streamwise velocity at the sphere wake of $x/D = 1.6$ and $2.5$. Note that the velocity is averaged within $0<\theta<2\pi$; (b) the angular distribution of the mean pressure coefficient around the sphere. The inset zooms more details.}
	\label{fig:mesh_convergence}
\end{figure}
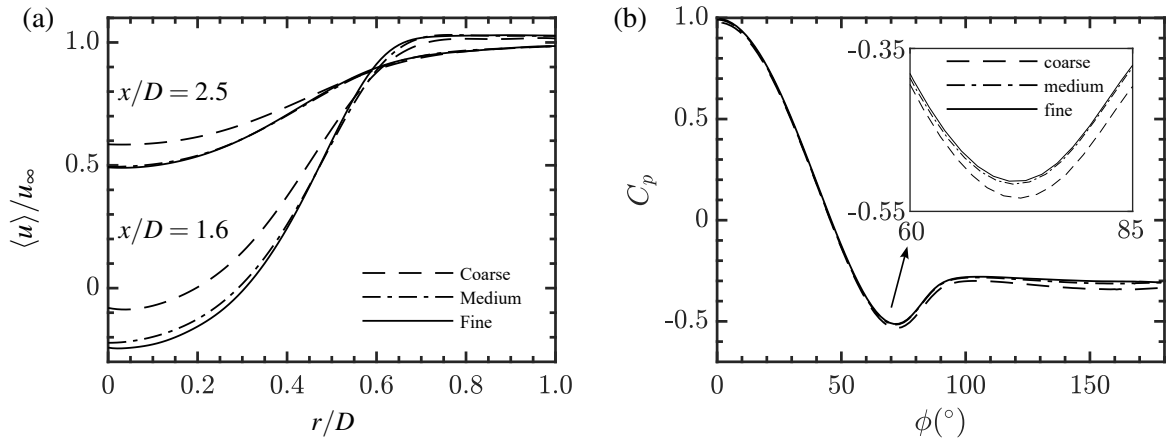

The section investigates the mesh convergence of the large-eddy simulation of the flow past a sphere at a Reynolds number of 10\,000 for Case-hexa. This is done by performing three simulations using different numbers of grid points based on the computational domain configuration shown in Fig.~\ref{fig:computational_domain}. Three distinct mesh resolutions are presented in Table~\ref{tab:mesh_convergence}, namely coarse (2\,539\,586), medium (5\,270\,032) and fine (10\,487\,744). These resolutions have been designed to achieve varying levels of refinement in the near-wall and wake regions. Note that the value of $y^+$, based on the surface friction velocity, is less than 1 for all three grid resolutions, which is sufficient to simulate the near-wall flow dynamics in LES. 

\begin{table}
	\begin{center}
    \caption{Simulation configurations and typical flow statistics for the coarse, medium, and fine meshes for flow past a sphere at $Re = 10\,000$.}
		\begin{ruledtabular}
		\begin{tabular}{lccccc}
			Simulations & Grid (total number) & $\Delta h_{min}/D$ & $y^{+}$ & $C_{D}$ & $\phi_s^{\circ}$  \\ [3pt]
            \hline
			Coarse & 2\,539\,586  & 0.00126  &  0.88  & 0.415 & 86.45 \\
			Medium & 5\,270\,032  & 0.0010    & 0.70 & 0.401 & 86.56 \\
			Fine   & 10\,487\,744 & 0.0008   & 0.55 & 0.397 & 86.52 \\
		\end{tabular}
        \end{ruledtabular}
		\label{tab:mesh_convergence}
	\end{center}
\end{table}

To evaluate whether the near-wall mesh is generated appropriately, typical flow parameters such as the mean drag coefficient and the boundary layer separation angle are compared in the three simulations in Table~\ref{tab:mesh_convergence}. Note that previous results used as reference are shown in Table~\ref{tab:Mean_properties}, and better agreement is observed for finer mesh resolutions. Furthermore, typical physical quantities, such as the time-averaged streamwise velocity and the mean pressure coefficient along the sphere in the three simulations, are compared in Fig.~\ref{fig:mesh_convergence}. The present results confirm that both quantities converge for the medium mesh used in this work to simulate flow past a sphere at $Re = 10\,000$.

\begin{figure}[!htpb]
	\centering
	\begin{minipage}[c]{1\linewidth}
		\centering
		\begin{overpic}[width=8cm]{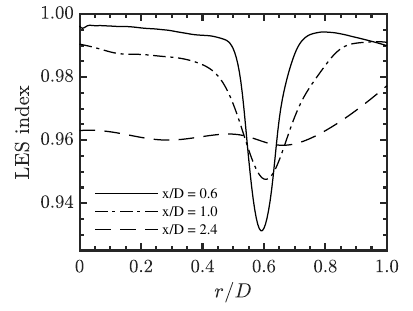}
		\end{overpic}
	\end{minipage}
	\caption[]{LES index at $0.6D$, $1.0D$, and $2.4D$ downstream of the sphere, and data are averaged along the azimuthal direction for medium grids (Table~\ref{tab:mesh_convergence}).}
	\label{fig:LES_index_figure}
\end{figure}

\begin{figure}[!htp]
  \centering
  \begin{minipage}[c]{1\linewidth}
  \centering
  \begin{overpic}[width=8cm]{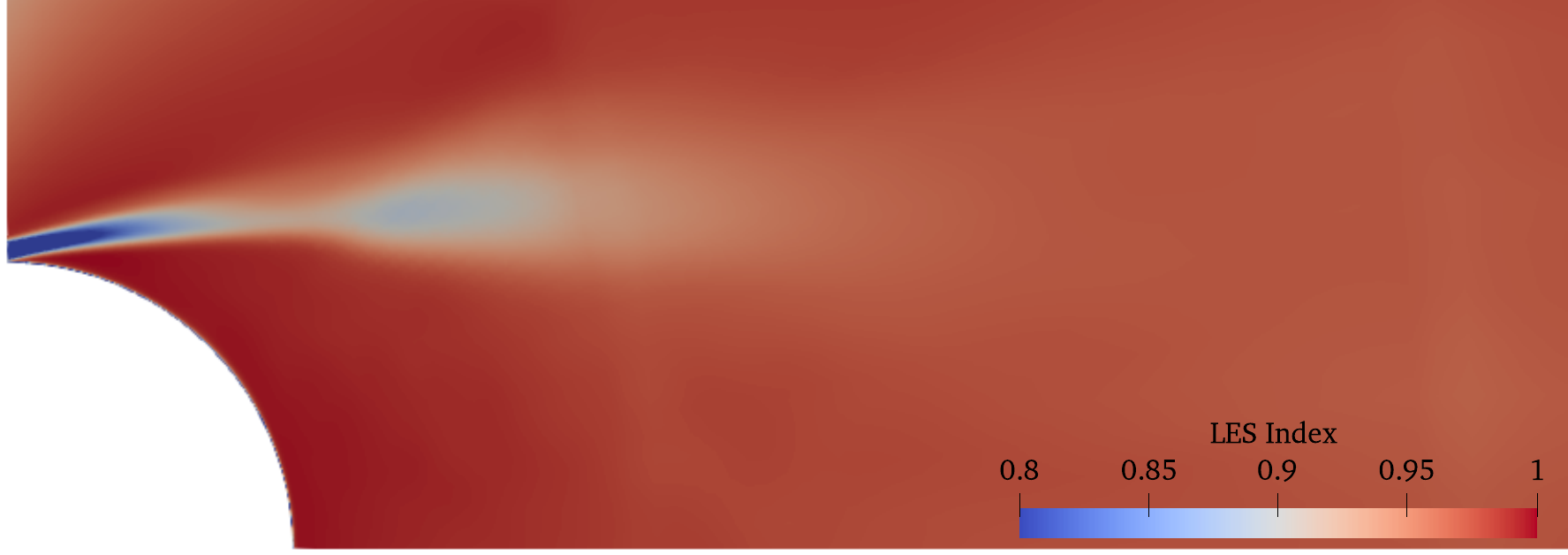}
  \put(2,30){\color{white}{(a)}}
  \end{overpic}
  \begin{overpic}[width=8cm]{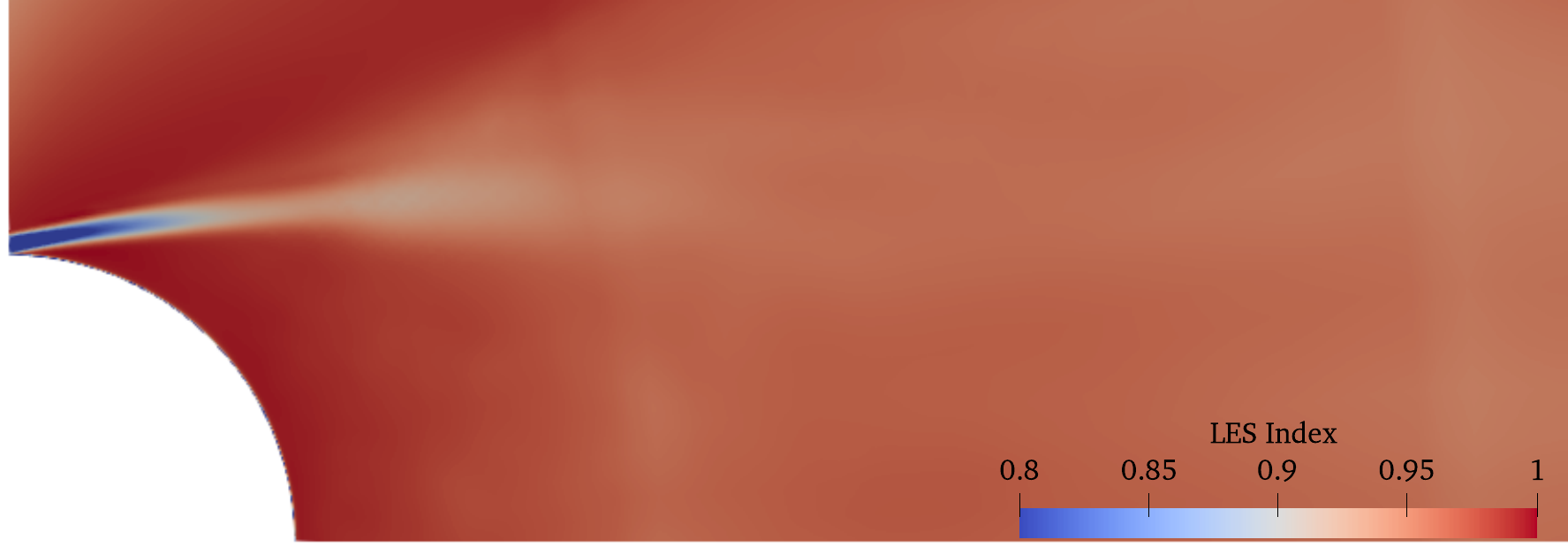}
  \put(2,30){\color{white}{(b)}}
  \end{overpic}
  \end{minipage}
  \caption[]{Snapshots of the ratio of resolved turbulent kinetic energy to total turbulent kinetic energy (Represent the LES Quality Index) (a) FVMS3, (b) pisoFoam. (Case-hexa, see Table~\ref{tab:Mesh Information})}
  \label{fig:LES_index_contour}
\end{figure}

In addition, the ratio of resolved to total TKE technique, defined as $\mathrm{{LES\ index}} =k_\mathrm{res}/(k_\mathrm{res}+k_\mathrm{sgs})$ (with the subscripts `$\mathrm{res}$' and `$\mathrm{sgs}$' representing the resolved and modeled TKE, respectively) has been used to check whether the current mesh resolution is appropriate for the LES. Previous studies suggest that the fraction of resolved TKE should be greater than 80\% for a qualified LES mesh, as shown in Pope \cite{Pope2000}. As shown in Fig.~\ref{fig:LES_index_figure}, we show the relative contributions of the resolved TKE to the total TKE at $0.6D$, $1.0D$, and $2.4D$ downstream of the sphere for the medium grid case. The ratio of resolved to total TKE predicted by the present solver is more than 90\% in most of the regions across the turbulent wake, which shows that the resolved TKE dominates the total TKE in the near-wake recirculation region for the present LES simulation. Therefore, the medium resolution is appropriate to simulate the flow past a sphere at $Re = 10\,000$.
\bibliographystyle{unsrt}
\bibliography{Main.bbl}
\end{document}